\documentclass[letterpaper,twocolumn,10pt]{article}
\usepackage{usenix}
\usepackage{fancyhdr}
\usepackage{epsfig,endnotes}
\usepackage{amsmath,amssymb}
\usepackage{multirow}
\usepackage{balance}
\usepackage[english]{babel}
\usepackage{algorithm,algpseudocode}
\usepackage{graphicx}
\usepackage{tablefootnote}
\usepackage{tikz}
\usepackage{url}
\usepackage{color}
\usepackage[export]{adjustbox}
\usepackage{mathptmx}
\usepackage{times}
\usepackage{hyperref}
\usepackage[hyphenbreaks]{breakurl}
\usepackage{subcaption}
\usepackage{booktabs}
\usepackage{longtable}

\usepackage{caption}
\usepackage{subcaption}

\renewcommand{\headrulewidth}{0.4pt}
\renewcommand{\footrulewidth}{0.4pt}

\newcommand{\para}[1]{{\vspace{1.5pt} \noindent \textbf{#1} \hspace{6pt}}}

\newcommand{\emed}[1]{{\color{black} #1}}

\newcommand{\shawnedit}[1]{{\color{black} #1}}
\newcommand{\revise}[1]{{\color{black} #1}}
\newcommand{\dave}[1]{{\color{black} #1}}
\newcommand{\htedit}[1]{{\color{black} #1}}

\newcommand{\secspace}{\vspace{0.0in}}

\newcommand{\pubfig}{{\tt PubFig}}
\newcommand{\facescrub}{{\tt FaceScrub}}
\newcommand{\webface}{{\tt WebFace}}
\newcommand{\vggfaceb}{{\tt VGGFace2}}

\newcommand{\vin}{{\tt VGG2-Incept}}
\newcommand{\vde}{{\tt VGG2-Dense}}
\newcommand{\win}{{\tt Web-Incept}}
\newcommand{\wde}{{\tt Web-Dense}}

\newcommand{\eg}{{\em e.g.,\ }}
\newcommand{\ie}{{\em i.e.\ }}
\newcommand{\etal}{{\em et al.\ }}

\newenvironment{packed_itemize}{
\begin{list}{\labelitemi}{\leftmargin=1.em}
  \setlength{\itemsep}{2pt}
  \setlength{\parskip}{0pt}
  \setlength{\parsep}{0pt}
  \setlength{\topsep}{0pt}
  \setlength{\partopsep}{0pt}
}{\end{list}}

\newfont{\mycrnotice}{ptmr8t at 7pt}
\newfont{\myconfname}{ptmri8t at 7pt}

\begin{document}

\title{Fawkes: Protecting Privacy against Unauthorized Deep Learning Models}
\author{Shawn Shan$^\dag$, Emily Wenger$^\dag$, Jiayun Zhang, Huiying Li, Haitao Zheng, Ben Y. Zhao\\
$^\dag$ denotes co-first authors with equal contribution\\
{\em Computer Science, University of Chicago}\\
{\em \{shansixiong, ewillson, huiyingli, htzheng, ravenben\}@cs.uchicago.edu, jiayunzhang15@outlook.com}}

\maketitle

\thispagestyle{plain}

\begin{abstract}

  Today's proliferation of powerful facial recognition \dave{systems} poses a real
  threat to personal privacy. As {\em Clearview.ai} demonstrated, anyone can canvas
  the Internet for data and train highly accurate facial recognition models
  of individuals without their knowledge. We need tools to protect ourselves from
  \dave{potential misuses of unauthorized facial recognition
  systems}. Unfortunately, \dave{no practical or effective solutions exist.}

  In this paper, we propose {\em Fawkes}, a system that \revise{helps individuals
  inoculate their} \dave{images} against unauthorized facial recognition models.
  Fawkes achieves this by helping users add imperceptible pixel-level
  changes (we call them ``cloaks'') to their own photos before
  \dave{releasing them}. When used to train
  facial recognition models, these ``cloaked'' images produce
  functional models that consistently \dave{cause normal images of the
    user to be misidentified}.  We
  experimentally demonstrate that Fawkes provides 95+\% protection against user
  recognition regardless of how trackers train their models. Even when clean,
  uncloaked images are ``leaked'' to the tracker and used for training, Fawkes
  can still maintain an 80+\% protection success rate.  \dave{We
    achieve 100\% success} in
  experiments against today's state-of-the-art facial recognition 
  services. Finally, we show that Fawkes is robust
  against a variety of countermeasures that try to detect or disrupt image cloaks.
\end{abstract}

\secspace
\section{Introduction}
\label{sec:intro}



Today's proliferation of powerful facial recognition models poses a real
threat to personal privacy. Facial recognition systems are scanning millions
of citizens in both the UK and China without explicit
consent~\cite{china-face,uk-face}. By next year, 100\% of international
travelers will be required to submit to facial recognition systems in top-20
US airports~\cite{airport-face}. Perhaps more importantly, anyone with
moderate resources can now canvas the Internet and build highly accurate
facial recognition models of us without our knowledge or awareness, {\em
  e.g.}  MegaFace~\cite{hill_krolik_2019}. Kashmir Hill from
the New York Times \dave{recently} reported on {\em Clearview.ai}, a private company that 
collected more than 3 billion online photos and trained a massive model capable
of recognizing millions of citizens, all without knowledge or
consent~\cite{clearview}.


\dave{Opportunities for} misuse of this technology are numerous and \dave{potentially} disastrous.
Anywhere we go, we can be identified at any time through street cameras,
video doorbells, security cameras, and personal cellphones. Stalkers can find
out our identity and social media profiles with a single
snapshot~\cite{stalker-face}. Stores can associate our precise in-store
shopping behavior with online ads and browsing
profiles~\cite{brazil-face}.  Identity thieves can easily identify (and
perhaps gain access to) our personal accounts~\cite{identity-face}.



\dave{We believe that private citizens need tools to protect themselves from being
  identified by unauthorized facial recognition models.}
Unfortunately, \dave{previous} work in this space is sparse and limited in both practicality
\htedit{and efficacy}. Some have
proposed distorting \emed{images to make them} unrecognizable and thus
avoiding facial
recognition~\cite{wu2018privacy, li2019anonymousnet, sun2018hybrid}. Others
produce adversarial patches in the form of bright patterns printed on
sweatshirts or signs,  which prevent facial recognition algorithms from even registering their
wearer as \emed{a person}~\cite{wu2019making,
  thys2019fooling}. Finally, given access to an image classification
model, ``clean-label poison attacks'' can \emed{cause the model} to
misidentify a single, \htedit{preselected} image~\cite{shafahi2018poison, zhu2019transferable}.

Instead, we propose {\em Fawkes}, a system that \revise{helps individuals to
inoculate their images} against unauthorized facial recognition models
\htedit{at any time without significantly distorting their own photos, or
  wearing conspicuous patches. Fawkes achieves this by helping users}  
adding imperceptible pixel-level changes (``cloaks'') to their own
photos. \htedit{For example, a} 
user who wants to share content ({\em e.g.} photos) on social
media or the public web can add small, imperceptible alterations to
\dave{their} photos before uploading them. If collected by a third-party ``tracker'' and
used to train a facial recognition model to recognize \dave{the user}, these
``cloaked'' images would produce functional models that consistently misidentify \dave{them}.

Our distortion or ``cloaking'' algorithm takes \dave{the user's} photos and computes
minimal perturbations that shift them significantly in the feature space of a
facial recognition model (using real or synthetic images of a third party as a
landmark). Any facial recognition model trained using these images of
\dave{the user} learns an altered set of ``features'' of what makes
\dave{them look like them.}
When presented with a clean, uncloaked image of \dave{the user}, {\em e.g.} photos from
a camera phone or streetlight camera, the model finds no labels associated
with \dave{the user} \htedit{in} the feature space near the image, and
classifies the photo \htedit{to another} 
label (identity) nearby in the feature space.

Our exploration of Fawkes produces several key findings:
\begin{packed_itemize} 
  \item We can produce significant alterations to images' feature
    space representations using \emed{perturbations imperceptible to} the naked eye (DSSIM $\leq 0.007$).
  \item Regardless of how the tracker trains its model (via transfer learning
    or from scratch), image cloaking provides 95+\%
    protection against user recognition (adversarial training techniques help
    ensure cloaks transfer to tracker models).
  \item \dave{Experiments show} 100\% success against state-of-the-art facial
    recognition services from Microsoft (Azure Face API), Amazon
    (Rekognition), and Face++. We first ``share'' our own (cloaked) photos as
    training data to each service, then apply the resulting models to
    \htedit{uncloaked} test images of the same person.
  \item In challenging scenarios where clean, uncloaked images are ``leaked'' to the
    tracker and used for training, we show how a single {\em Sybil} identity
    can boost privacy protection. This results in 80+\% success \htedit{in
   avoiding identification} even when half of the training images are uncloaked.
  \item Finally, we consider a tracker who is aware of our image cloaking
    techniques and evaluate the efficacy of potential countermeasures. We
    show that image cloaks are robust (maintain high protection rates
    against) to a variety of mechanisms for both cloak disruption and cloak
    detection.
\end{packed_itemize}

\secspace


\secspace
\section{Background and Related Work}
\label{sec:back}


To protect user privacy, our image cloaking techniques leverage and extend
work broadly defined as poisoning attacks in machine learning. Here, we
\emed{set the} context by discussing prior \emed{efforts to} help
users evade facial recognition models. We then discuss relevant data
poisoning attacks, followed by related work on privacy-preserving machine learning and \htedit{techniques to train
facial recognition models.}


Note that to protect user privacy from unauthorized deep
learning models, we employ attacks against ML models. In this
scenario, {\em users} are the ``attackers,'' and third-party {\em trackers} running unauthorized tracking are the ``targets.'' 

\secspace
\subsection{Protecting Privacy via Evasion Attacks}
Privacy advocates have considered the problem of protecting individuals from
facial recognition systems, generally by making images difficult for a facial
recognition model to recognize. Some rely on creating {\em adversarial
  examples}, inputs to the model designed to cause 
misclassification~\cite{szegedy2013intriguing}.
These attacks have since been proven possible ``in the wild,''
Sharif~\etal~\cite{sharif2016accessorize} create specially printed glasses 
that cause the wearer to be misidentified. Komkov and
Petiushko~\cite{komkov2019advhat} showed that carefully computed adversarial
stickers on a hat can reduce \emed{its wearer's likelihood of being
  recognized}.  Others propose ``adversarial patches'' that target ``person
identification'' models, making it difficult for \emed{models to recognize
  the wearer as a person} in an image~\cite{wu2019making,thys2019fooling}.

All of these approaches share two limitations. First, they require the user
to wear fairly obvious and conspicuous accessories (hats, glasses, sweaters)
that are impractical for normal use. Second, in order to evade tracking,
they require {\em full and unrestricted} access (white box access) to the precise model
tracking them. Thus they are easily broken (and user privacy compromised) by
any tracker that updates its model.

Another line of work seeks to {\em edit facial images} so
that human-like characteristics are preserved but facial recognition
model accuracy is significantly reduced. Methods used include
k-means facial averaging~\cite{newton2005preserving},
facial inpainting~\cite{sun2018natural}, and GAN-based face
editing~\cite{wu2018privacy,li2019anonymousnet, sun2018hybrid}. Since these
dramatically alter the user's face in her photos, we consider them
impractical for protecting shared content.

\begin{figure*}[t]
  \centering
  \includegraphics[width=0.9\linewidth]{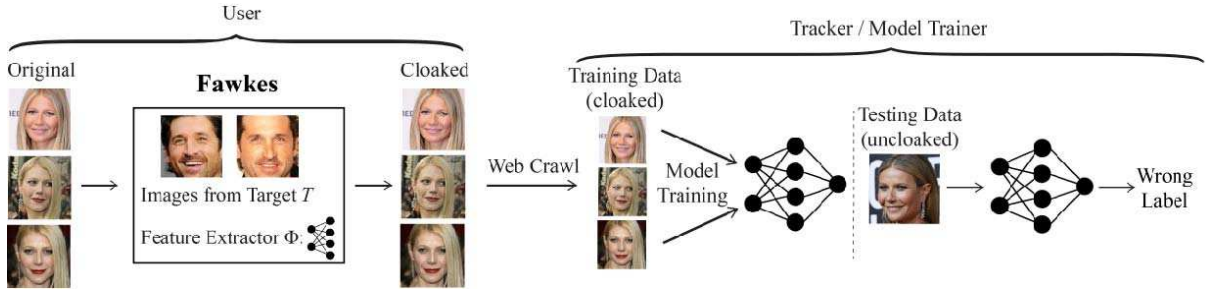}
  \caption{Our proposed Fawkes system that protects user privacy by
    cloaking their online photos. (Left)  A user $U$ applies cloaking algorithm (given
    a feature extractor $\Phi$ and images from some target $T$) to generate
    cloaked versions of $U$'s photos, each with a small perturbation
    unnoticeable to the human eye.  (Right) A tracker crawls the cloaked images from
    online sources, and uses them to train an (unauthorized) model to recognize and track
    $U$. When it comes to classifying new (uncloaked) images of $U$, the
    tracker's model misclassifies them to someone not $U$.  Note that
    $T$ does not have to exist in the tracker's model. }
  \label{fig:system_overview}
\end{figure*}

\secspace
\subsection{Protecting Privacy via Poisoning Attacks}
An alternative to evading models is to disrupt their training.
This approach leverages ``data poisoning attacks'' against deep learning 
models. These attacks \htedit{affect} deep learning models by modifying
the initial data used to train them, usually by adding a set of samples $S$
and associated labels $L_S$. 
Previous work has used data poisoning to induce unexpected behaviors
in trained DNNs~\cite{yang2017generative}. In this section, we discuss
two data poisoning attacks related to our work, and
\shawnedit{identify their key limitations when used to protect user privacy. }

\para{Clean Label Attacks.}  A clean-label poisoning attack injects
``correctly'' labeled poison images into training data, causing a model
\emed{trained on this data} to misclassify a specific image of
interest~\cite{shafahi2018poison,zhu2019transferable}. What distinguishes
clean-label attacks from normal poisoning attacks is that all image labels
remain unchanged during the poisoning process -- only the content of the
poisoned images changes.

Our work (Fawkes) works with similar constraints. Our \htedit{action} to affect
or disrupt a model is limited to altering a group of images with a correct
label, {\em i.e.} a user can alter her images but \emed{cannot claim
  these   are images of someone else.}

Current clean label attacks cannot address the privacy problem because of
three factors. {\em First}, they only cause misclassification on a {\em
  single, preselected} image, whereas user privacy protection requires the
misclassification of any current or future image of the protected user
({\em i.e.\/} an
entire model class).  {\em Second}, clean label attacks do not transfer well
to different models, especially models trained from scratch. Even between
models trained on the same data, the attack only transfers with $30\%$
success rate~\cite{zhu2019transferable}.  {\em Third}, clean label attacks
are easily detectable through anomaly detection in the feature
space~\cite{gupta2019strong}. 




\para{Model Corruption Attacks.} Other recent work proposes techniques to
modify images such that they {\em degrade} the accuracy of a model trained on
them~\cite{shen2019tensorclog}. The goal is to spread these poisoned images
in order to discourage unauthorized data collection and model training. We
note that Fawkes' goals are to mislead rather than frustrate. Simply
corrupting \htedit{data} of a user's class may inadvertently inform the tracker of the user's
evasion attempts and lead to more advanced countermeasures by the tracker.
\emed{Finally,~\cite{shen2019tensorclog} only has a 50\% success rate in
  protecting a user from being recognized.}



\secspace
\subsection{Other Related Work}
\para{Privacy-Preserving Machine Learning.}
Recent work has shown that ML
models can memorize (and subsequently leak) parts of their training
data~\cite{song2017machine}. This can be exploited to expose private details
about members of the training dataset~\cite{fredrikson2015model}. These
attacks have spurred a push towards {\em differentially private} model
training~\cite{abadi2016deep}, which uses techniques from the field of
differential privacy~\cite{dwork2008differential} to protect sensitive
characteristics of training data. We note these techniques imply a trusted
model trainer and are ineffective against an unauthorized model trainer.

\para{Feature Extractors \& Transfer Learning.}
\label{sec:translearn}
Transfer learning uses existing
pretrained models as a basis for quickly training models for customized
classification tasks,  using less training data. Today, it is commonly
used to deploy complex ML models ({\em e.g.\/} facial
recognition or image segmentation~\cite{transfer2014}) at reasonable
training costs. 

In transfer learning, the knowledge of a pre-trained feature extractor
$\Phi$ is passed on to a new model $\mathbb{F}_{\theta}$. Typically, a
model $\mathbb{F}_{\theta}$ can be created by
appending a few additional layers to $\Phi$ and only training 
those new layers. The original layers that composed $\Phi$ will remain
unmodified. As such, pre-existing knowledge ``learned'' by $\Phi$ is
passed on to the model 
$\mathbb{F}_{\theta}$ and directly influences its 
classification outcomes.  Finally,  transfer learning is most
effective when the feature extractor and model
are trained on similar datasets. For example, a facial recognition
model trained on faces extracted from YouTube videos might serve well
as a feature extractor for a model designed to recognize celebrities in magazines.

Finally, the concept of protecting individual privacy against invasive
technologies extends beyond the image domain. Recent work~\cite{jammer} proposes
wearable devices that restore personal agency using digital jammers to
prevent audio eavesdropping by ubiquitous digital home assistants.

\secspace
\section{Protecting Privacy via Cloaking}
\label{sec:method}

\secspace

We propose {\em Fawkes}, a system \revise{designed to help protect the
  privacy of a {\em user}} against unauthorized facial recognition models
trained by a third-party {\em tracker} on the user's images. Fawkes
achieves this by adding subtle perturbations (``cloaks'') to the
user's images before sharing them.  Facial recognition models trained on
cloaked images will have a distorted view of the user in the ``feature
space,'' {\em i.e.} the model's internal understanding of what makes the user
unique. Thus the models cannot recognize real (uncloaked) images of the user,
and instead, misclassify them as someone else.

In this section, we first describe the threat model and assumptions
for both users and trackers. \htedit{We then present the intuition behind
cloaking and our methodology to generate cloaks. Finally, we discuss
why cloaking by individuals is effective against unauthorized facial
recognition models.} 


\secspace
\subsection{Assumptions and Threat Model}
\label{sec:assumption}

\para{User.} The user's goal is to share their photos online without
unknowingly helping third party trackers build facial recognition models that
can recognize them.  \htedit{Users protect themselves by adding imperceptible
  perturbations (``cloaks'') to their photos before sharing them. This is
  illustrated in the left part of Figure~\ref{fig:system_overview}, where a
  cloak is added to this user's photos before they are uploaded.}

The design goals for these cloaks are:
\begin{packed_itemize} \vspace{-0.06in}
\item cloaks should be \textbf{imperceptible} and not impact normal use of
  the image;
\item \htedit{when classifying normal, uncloaked images}, models trained on
  cloaked images should recognize the underlying person with \textbf{low
    accuracy}. \vspace{-0.06in}
\end{packed_itemize}
We assume the user has access to moderate computing resources
(\eg a personal laptop) and applies cloaking to their own images locally.  We
also assume the user has access to some feature extractor, {\em e.g.} a
generic facial recognition 
model, represented as $\Phi$ in Figure~\ref{fig:system_overview}.  
Cloaking is simplified if the user has the same $\Phi$ as the
tracker.  \htedit{We begin with this common assumption (also used by
  prior work~\cite{wang2018great,shafahi2018poison,zhu2019transferable}),
  since only a few large-scale face recognition models are available in
the wild.}  Later in \S\ref{subsec:transferability}, we relax this assumption and show how
our design maintains the above properties. 



We initially consider the case where the user has the ability to apply
cloaking to all their photos to be shared, thus the tracker can only collect cloaked
photos of the user.  Later in \S\ref{sec:sybil}, we explore a scenario where
a stronger tracker has obtained access to some number of their uncloaked
images.

\para{Tracker/Model Trainer.} We assume that the tracker (the entity
training unauthorized models) is a third party without direct access to user's
personal photos ({\em i.e.} not Facebook or
Flickr). The tracker could be a company like Clearview.ai, a
government entity, or even an individual. The tracker has significant
computational resources. They can either use transfer learning to
simplify their model training process (leveraging existing feature
extractors), or train their model completely from scratch. 


We also assume the tracker's primary goal is to build a powerful model to track
many users rather than targeting a single specific person\footnote{Tracking a
  specific person can be easily accomplished through easier, offline methods,
  {\em e.g.} a private investigator who follows the target user, and is beyond the
  scope of our work.}.  The tracker's primary data source is a collection
of public images of users obtained via web scraping. We also consider scenarios where
they are able to obtain some number of uncloaked images from other sources
(\S\ref{sec:sybil}).


\revise{\para{Real World Limitations.} Privacy benefits of Fawkes rely on
  users applying our cloaking technique to the majority of images
  \dave{of their likeness} before posting online.  In practice,
  however, users are unlikely to control all images of themselves, such as photos shared online
  by friends and family, media, employer or government websites. While it is
  unclear how easy or challenging it will be for trackers to associate these
  images \dave{with} the identity of the user, a tracker who obtains a large number of
  uncloaked images of the user can compromise the effectiveness
  of Fawkes.

  Therefore, Fawkes is most effective when used in conjunction with other
  privacy-enhancing steps that minimize the online availability of a
  user's uncloaked images. For example, users can curate their social media
  presence and remove tags of their names applied to group photos on Facebook
  or Instagram. Users can also leverage privacy laws such as ``Right to be
  Forgotten'' to remove and untag online content related to
  themselves. The online curation of personal images is a challenging
  problem, and we leave the study of minimizing online image footprints to
  future work.}

\secspace
\subsection{Overview and Intuition}
DNN models are trained to identify and extract (often hidden) {\em features}
in input data and use them to perform classification.
Yet their ability to identify features is easily disrupted
by data poisoning attacks during model training, where
small perturbations on training data with a particular label ($l$) can shift
the model's view of what features uniquely identify
$l$~\cite{shafahi2018poison, zhu2019transferable}. \emed{Our work
leverages this property to cause misclassification
of {\em any existing or future image} of a single class,
providing one solution to the challenging problem of protecting
personal privacy against the unchecked spread of facial recognition models.}




Intuitively, our goal is to \htedit{protect a user's privacy by modifying their photos} in small and imperceptible
ways before posting them online, such that a facial
recognition model trained on them learns the wrong features about what makes
 \dave{the user look like the user}. The model thinks it is successful,  because it
correctly recognizes its sample of (modified) images of the user. However, when unaltered images of \dave{the user}, {\em e.g.} from a surveillance video, are fed into the model, the model does not detect
the features it associates with \dave{the user}. Instead, it identifies someone else as
the person in the video. \htedit{By simply modifying their online
  photos, \dave{the user} successfully prevents unauthorized trackers and their
  DNN models from recognizing \dave{their} true face.} 

\secspace
\subsection{Computing Cloak Perturbations}
But how do we determine what perturbations (we call them ``cloaks'') to apply
to Alice's photos?  An effective cloak would
teach a face recognition model to associate
Alice with erroneous features that are quite different from real
features defining  Alice. Intuitively, the more dissimilar or distinct these erroneous features
are from the real Alice, the less likely the model will be able to recognize
the real Alice.

In the following, we describe our methodology for computing cloaks for
each specific user, with the goal of making the features
learned from cloaked photos highly dissimilar from those learned from
original (uncloaked) photos.

\para{Notation.}  Our discussion will use the following notations. 
\begin{packed_itemize} \vspace{-0.05in}
\item $x$:  Alice's image (uncloaked) 
\item $x_T$:   target image (image from another class/user $T$) used to generate cloak for Alice
\item $\delta (x, x_T)$: cloak computed for Alice's image $x$ based on
  an image $x_T$ from label $T$
\item $x \oplus \delta (x,x_T)$: cloaked version of Alice's image $x$
\item $\Phi$: Feature extractor used by facial recognition model 
\item  $\Phi(x)$: Feature vector (or feature representation) extracted from an input $x$
\end{packed_itemize}\vspace{-0.08in}


\para{Cloaking to Maximize Feature Deviation.}  Given each photo ($x$)
of Alice to be shared online,  our ideal cloaking design modifies $x$
 by adding a cloak perturbation $\delta (x, x_T)$ to $x$ that
 maximize changes in $x$'s feature representation: 
 \begin{eqnarray}
   &\max_{\delta} Dist\left(\Phi (x), \Phi(x \oplus \delta (x, x_T))\right),  \label{eq:cloakopt}\\
  & \text{subject to } \; |\delta (x, x_T)|< \rho, \nonumber
\end{eqnarray} 
where $Dist(.)$ computes the distance of two feature vectors,
$|\delta|$ \emed{measures} the perceptual perturbation caused by cloaking,
and $\rho$ is the perceptual perturbation budget.

To guide the search for the cloak perturbation in eq (\ref{eq:cloakopt}), 
we use another image $x_T$ from a different
user class ($T$).  Since the feature space $\Phi$ is highly complex,
$x_T$ serves as a landmark, enabling fast and efficient search for the
input perturbation that leads to large changes in feature
representation.  Ideally, $T$ should be very dissimilar from
Alice in the feature space. We illustrate this in 
Figure~\ref{fig:system_overview}, where we use Patrick
Dempsey (a male actress) as a dissimilar 
target $T$ for the original user (female actor Gwyneth Paltrow). 


We note that our design does not assume that the cloak target ($T$)
and \emed{the associated} $x_T$  are
used by any tracker's face recognition model. In fact, any user whose
feature representation is sufficiently different from Alice's would suffice
(see \S\ref{subsec:transferability}).    \emed{Alice can easily check
for such dissimilarity by} running the feature extractor $\Phi$ on
other users' online photos.  Later in \S\ref{sec:design} we will present the
detailed algorithm \emed{for} choosing the target user $T$ from
public datasets \emed{of facial images}. 

\para{Image-specific Cloaking.}  When creating cloaks for her photos,
Alice will produce image-specific cloaks, {\em i.e.\/} $\delta(x,x_T)$
is image dependent.  Specifically, Alice will pair each original image
$x$ with a target image $x_T$ of class $T$.  In our current
implementation, the search for  $\delta(x,x_T)$ replaces the ideal
optimization defined by eq. (\ref{eq:cloakopt}) with the following optimization: 
\begin{eqnarray}
   &\min_{\delta} Dist\left(\Phi (x_T), \Phi(x \oplus \delta (x, x_T))\right),  \label{eq:optstop}\\
  & \text{subject to } \; |\delta (x, x_T)|< \rho.  \nonumber
\end{eqnarray} 
Here we search for the cloak for $x$ that shifts its feature
representation closely towards $x_T$. 
This new form of optimization also prevents the system from generating extreme $\Phi(x \oplus
\delta (x, x_T))$ values that can be easily detected by trackers using
anomaly detection. 

Finally, our image-specific cloak optimization will create different cloak patterns
among Alice's images.  This ``diversity'' makes it hard for trackers
to detect and remove cloaks. 


\secspace
\subsection{Cloaking Effectiveness \& Transferability}
\label{subsec:transferability}

Now a user (Alice) can produce 
cloaked images whose feature representation is dissimilar
from \emed{her own} but similar to that of a target user $T$. But does
this translate into the desired misclassification behavior in the
tracker model?  Clearly, if $T$ is a class \emed{in the} tracker model,
Alice's original (uncloaked) images will not be classified as Alice.
But under the more likely scenario where $T$ is not in the tracker model,  does cloaking still
lead to misclassification?

We believe the answer is {\bf yes}.  Our
hypothesis is that as long as the feature representations of
Alice's cloaked and uncloaked images are sufficiently different, the
tracker's model will not classify them \emed{as} the
same class.  This is because there will be another user class ({\em
  e.g.\/} $B$) in the tracker model, whose feature representation is
more similar to  $\Phi(x)$ (true Alice) than
$\Phi(x\oplus\delta)$ (Alice learned by the model).  Thus, the
model will classify Alice's normal images as $B$.

\begin{figure}[t]
  \centering
  \includegraphics[width=1\columnwidth]{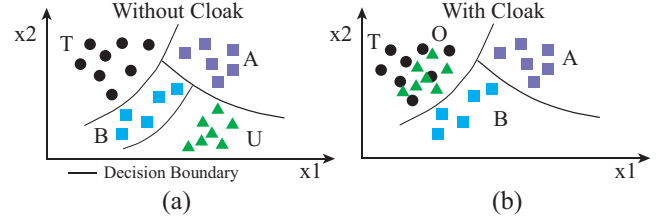}
  \caption{The intuition for why a \emed{tracker's model trained on
      $U$'s 
  cloaked photos will misclassify} $U$'s original photos,
    visualized on a simplified 2D feature space with four user classes
    $A$, $B$, $U$ (aka Alice), $T$. (a) decision
    boundaries of the model trained on $U$'s uncloaked
   photos. (b) decision boundaries when trained on
    $U$'s cloaked photos (with target $T$). }
 
  \label{fig:intuition}
\end{figure}

We illustrate this in  Figure~\ref{fig:intuition} using a simplified
2D visualization of the feature space. \emed{There are} 4 classes ($A$, $B$,
$U$ aka Alice, and $T$) \emed{that a tracker wishes to distinguish}. The two figures show the tracker 
model's decision boundary when $U$'s training data is uncloaked and
cloaked, respectively.  In Figure~\ref{fig:intuition}(a), the model
will learn $U$'s true feature representation as the bottom right corner.  In Figure~\ref{fig:intuition}(b), $U$ uses $T$
as the cloak target, and the resulting tracker model will learn $U$'s
feature representation $\Phi(x\oplus\delta)$ as green triangles near
$T$ (top left corner).  This means that the area corresponding to $U$'s original feature
representation $\Phi(x)$ will be classified as $B$.  More importantly,
this
(mis)classification 
will occur whether \emed{or not} $T$ is a class in the tracker's
model. 

Our above discussion assumes \emed{the tracker's model contains a
class} whose feature representation is more similar to the user's 
original feature representation than  her cloaked feature
representation.  This
is a reasonable assumption when the tracker's model targets many
users ({\em e.g.\/} 1,000) rather than a few users ({\em e.g.\/}
2). \htedit{Later in \S\ref{sec:eval} we confirm that cloaking
  is highly effective against multiple facial recognition models with
  anywhere from 65 to 10,575 classes. }

\para{Transferability.}  \htedit{Our above discussion also assumes that the user
has the same feature extractor $\Phi$ as is used to train the tracker model.   Under
the more general scenario, } the effectiveness of cloaking
\htedit{against 
any tracker models} relies on the {\em
transferability} effect, the property that models trained
for \emed{similar tasks} share similar properties and
vulnerabilities, even when they were trained on different architectures and different training
data~\cite{demontis2019adversarial,transfer,suciu2018does,transfer2014}.

\emed{This transferability property suggests that cloaking should still be
effective even if the tracker performs transfer learning using a different feature extractor or
trains their model from scratch. Because the user's and tracker's
feature extractors/models are designed for similar tasks ({\em i.e.} facial
recognition), cloaks should be effective regardless of the
tracker's training method.} 
\htedit{Later, we empirically evaluate cloaking success rate when trackers
use different feature extractors (\S\ref{sec:scenario2}) or train
models from scratch (\S\ref{sec:scenario3}). In all scenarios, cloaking is highly
effective ($> 95\%$ protection rate).}

\secspace
\section{The Fawkes Image Cloaking System}
\label{sec:design}
We now present the detailed design of
{\em Fawkes}, a practical image cloaking system
\htedit{that allows users to evade identification} by unauthorized
facial recognition models.  Fawkes \emed{uses} three 
steps \emed{to help a} user modify and publish her online photos. 

\emed{Given} a user $U$, Fawkes takes as input the set of $U$'s photos to
be shared online $\mathbf{X_U}$,  the (generic) feature extractor
$\Phi$, and the cloak perturbation budget $\rho$.

\para{Step 1: Choosing a Target Class $T$.} First, Fawkes examines a
publicly available dataset \emed{that} contains numerous groups of
images, each identified with a specific class label, {\em e.g.} Bob,
Carl, Diana. Fawkes randomly picks $K$
candidate target 
classes and their images from this public dataset and uses the feature
extractor $\Phi$ to calculate $\mathcal{C}_k$, the centroid of the feature space for
each class $k=1..K$.  Fawkes picks \emed{as} the target class $T$ the class in the $K$
candidate set whose feature representation centroid is most
dissimilar from \emed{the feature representations} of all images
in $\mathbf{X_U}$, {\em i.e.\/}  \begin{equation} \label{eq:target_selection} \vspace{-0.05in}
 T=  \underset{\boldsymbol{k=1..K}}{\text{argmax}} \underset{\boldsymbol{{x\in \mathbf{X_U}}}}{\text{min}}  \; Dist (\Phi(x),
 \mathcal{C}_k). 
\end{equation}
\emed{We use L2 as the distance function in feature space, $Dist(.)$.}



\para{Step 2: Computing Per-image Cloaks.}  Let $\mathbf{X_T}$
represent the set of target images available to user $U$. For each image of user $U$,
$x\in \mathbf{X_U}$, Fawkes randomly picks an image $x_T\in \mathbf{X_T}$, and computes a cloak $\delta(x,x_T)$ for $x$, following
the optimization defined by eq. (\ref{eq:optstop}), subject to
$|\delta(x,x_T)|<\rho$.  


In our implementation, $|\delta(x,x_T)|$ is calculated using the DSSIM (Structural Dis-Similarity Index)~\cite{ssim, msssim}.  Different from the $L_p$ distance used in previous work~\cite{carlini2017,kurakin2016adversarial,shan2019gotta}, DSSIM has
gained popularity as a measure of user-perceived image
distortion~\cite{wang2018great,jan2019connecting,li2019hiding}.  Bounding cloak generation
with this metric ensures that cloaked versions of images are visually similar
to the originals.




We apply the \textit{penalty
  method}~\cite{nocedal2006numerical} to reformat and solve the optimization
in eq.(\ref{eq:optstop}) as follows: 
\begin{equation*} \vspace{-0.05in}
 \underset{\delta}{\text{min }} Dist\left(\Phi (x_T), \Phi(x \oplus \delta (x,
   x_T))\right) + \lambda \cdot max (|\delta(x,x_T)|-\rho,0)
 \end{equation*}


 \noindent Here $\lambda$ controls the impact of the input
 perturbation caused by cloaking.  When $\lambda \rightarrow \infty$,
 the cloaked image is visually identical to the original image. Finally, to ensure the
 input pixel intensity remains in the correct range ($[0,255]$), we
 transform the intensity values into $tanh$ space as proposed in previous
work~\cite{carlini}.


\para{Step 3: Limiting Content.} Now the user $U$ has created the set
of cloaked images that she can post and share online.  However,
the user must be careful to ensure that no uncloaked images are
shared online and associated with her identity.  Any images shared by friends
and labeled or tagged with her name would provide uncloaked training data for
a tracker model. Fortunately, a user can proactively ``untag'' herself
on most photo sharing sites.

Even so, a third party might be able to restore
those labels and re-identify her in those photos using friendlist
intersection attacks~\cite{osn-intersect}. Thus,
in~\S\ref{sec:sybil}, we expand the design of Fawkes to address
trackers who are able to obtain uncloaked images in addition to
cloaked images of the user.

\secspace
\section{System Evaluation}
\label{sec:eval}

\begin{table*}[t]
  \centering
  \resizebox{0.8\textwidth}{!}{
    \begin{tabular}{|c|c|c|c|c|c|c|}
      \hline
      \multicolumn{1}{|c|}{\multirow{2}{*}{\textbf{Teacher Dataset}}}
      & \multicolumn{1}{c|}{\multirow{2}{*}{\textbf{Model  Architecture}}}
      & \multicolumn{1}{c|}{\multirow{2}{*}{\textbf{Abbreviation}}}
      &
      \multicolumn{1}{c|}{\multirow{2}{*}{\textbf{\begin{tabular}[c]{@{}c@{}}Teacher
              Testing \\ Accuracy\end{tabular}}}}
      &
      \multicolumn{2}{c|}{\textbf{Student Testing Accuracy}}
      \\ \cline{5-6} \multicolumn{1}{|c|}{}
      & \multicolumn{1}{c|}{}
      & \multicolumn{1}{c|}{} &   \multicolumn{1}{c|}{}
      & \multicolumn{1}{c|}{\textbf{PubFig  }}                 & \multicolumn{1}{c|}{\textbf{FaceScrub}} \\ \hline
     \webface{}
      & InceptionResNet
      & \win{}  &$74\%$   & $96\%$  & $92\%$                               \\ \hline
      \webface{}                                                       & DenseNet                          & \wde{}                               & $76\%$                                                                & $96\%$                                              & $94\%$                               \\ \hline
      \vggfaceb{}
      & InceptionResNet                        & \vin{}                           & $81\%$                                                                & $95\%$                                              & $90\%$                               \\ \hline
      \vggfaceb{}
      & DenseNet                                            & \vde              & $82\%$                                                                & $96\%$                                              & $92\%$                               \\ \hline
    \end{tabular}
  }
  \caption{The  four feature extractors used in our evaluation, their classification efficacy and those of
    their student models.}
    \label{tab:model_eval}  
\end{table*}

\begin{table}[t]
  \centering
  \resizebox{0.45\textwidth}{!}{
    \begin{tabular}{|c|r|c|r|}
      \hline
      \textbf{Dataset} & \textbf{\# of Labels} & \textbf{Input Size}       & \textbf{\# of Training Images} \\ \hline
      \pubfig{}        & $65$                  & $224 \times 224 \times 3$ & $5,850$                        \\ \hline
      \facescrub{}     & $344$                 & $224 \times 224 \times 3$ & $37,905$                       \\ \hline
      \webface{}          & $10,575$              & $224 \times 224 \times 3$ & $475,137$                      \\ \hline
      \vggfaceb{}      & $8,631$               & $224 \times 224 \times 3$ & $3,141,890$                    \\ \hline
    \end{tabular}

  }
  \caption{Datasets emulating user images in experiments.}
  \label{tab:task_detail}   
\end{table}

In this section, we evaluate the effectiveness of 
Fawkes.  We first describe the datasets, models, and experimental
configurations used in our tests. We then present results for cloaking in three
different scenarios: 1) the user produces cloaks using the same feature
extractor as the tracker; 2) the user and tracker use different feature extractors; and 3) the tracker trains models from
scratch (no feature
extractor). 

Our key findings are: cloaking is highly effective when users share a feature
extractor with the tracker; efficacy could drop when feature extractors are
different, but can be restored to near perfection by making the
user's feature extractor robust (via adversarial training); and,
similarly, cloaks generated on robust feature extractors work well
even when trackers train models from scratch. 

\secspace
\subsection{Experiment Setup}
Our experiments require two components. First, we need feature extractors
that form the basis of facial recognition models for both \emed{the} user's cloaking purposes
and the tracker's model training. Second, we need datasets that
emulate a set of user images scraped by the tracker and enable us to evaluate the impact of cloaking.

\para{Feature Extractors.} There are few publically available,
large-scale facial recognition models. Thus we train feature extractors using two large
($\ge$ $500$K images) datasets on different model architectures
(details in Table~\ref{tab:task_detail}). 

\begin{packed_itemize}\vspace{-0.06in}
\item \vggfaceb{} contains $3.14$M images of $8,631$ subjects downloaded
  from Google Image Search~\cite{cao2018vggface2}.
\item \webface{} has $500,000$ images of faces covering roughly $10,000$
  subjects collected from the Internet~\cite{yi2014learning}. \vspace{-0.06in}
\end{packed_itemize}

Using these two datasets, we build four feature extractors, two from
each. We use two different model 
architectures: a) DenseNet-121~\cite{huang2017densely}, a $121$ layer neural
network with $7$M parameters, and b) InceptionResNet
V2~\cite{szegedy2017inception}, a $572$ layer deep neural network with over
$54$M parameters. Our trained models have comparable accuracy with
previous work~\cite{cao2018vggface2, wang2018great, nech2017level} and
perform well in transfer learning scenarios.  For clarity, we abbreviate feature
extractors based on their dataset/architecture
pair. Table~\ref{tab:model_eval} lists \emed{the classification accuracy for our feature extractors and student models.}

\para{Tracker's Training Datasets.}  Under the scenario where the
tracker trains its facial recognition model from scratch (\S\ref{sec:scenario3}), we assume
they will use the above two large datasets (\vggfaceb{}, \webface{}).
Under the scenario where they apply transfer learning
(\S\ref{sec:scenario1} and \S\ref{sec:scenario2}), the tracker uses
the following two smaller datasets (more details in Table~\ref{tab:task_detail}).  


\begin{packed_itemize} \vspace{-0.06in}
\item \pubfig{} contains $5,850$ training images and $650$
  testing images of $65$ public figures\footnote{We exclude $18$ celebrities
    also used in the feature extractor datasets.}~\cite{pubfig}.
\item \facescrub{} contains $100,000$ images of $530$ public
  figures on the Internet~\cite{ng2014data}\footnote{We
    could only download $60,882$ images for $530$ people, as some
    URLs were removed. Similarly, prior work~\cite{auxccs} only 
    retrieved $48,579$ images.}. \vspace{-0.06in}
\end{packed_itemize}
To perform
transfer learning, the tracker adds a softmax layer at the end of the
feature extractor (see~\S\ref{sec:translearn}), and fine-tunes the
added layer using the above dataset.

\para{Cloaking Configuration.}  In our experiments, we randomly choose
a user class $U$ in the tracker's model, {\em e.g.\/} a random user in
\pubfig{}, to be the user seeking protection.   We then apply the target selection algorithm described in
\S\ref{sec:design} to select a target class $T$ from a small subset of
users in \vggfaceb{} and \webface{}. Here we ensure that $T$ is not a
user class in the tracker's model.

For each given $U$ and $T$ pair, we pair each image $x$ of $U$ with an
image $x_T$ from $T$, and compute the cloak for $x$.  For this we run the Adam optimizer for $1000$ iterations with a learning rate
of $0.5$. 

As discussed earlier, we evaluate our cloaking under three scenarios,
$U$ and tracker model sharing the same feature extractor
(\S\ref{sec:scenario1}),  the two using
different feature extractors (\S\ref{sec:scenario2}),  and the tracker training model from
scratch without using any pre-defined feature extractor
(\S\ref{sec:scenario3}).




\para{Evaluation Metrics.} In each scenario, we evaluate
cloak performance using two metrics: {\em protection success rate},
which is the tracker model's 
misclassification rate for clean (uncloaked) images of $U$,  and
{\em normal accuracy}, which is the overall classification accuracy of the
tracker's model on users beside $U$.   When needed, we
indicate the configuration of user/tracker feature extractors using
the notation <entity>:<feature extractor>. 


\begin{figure*}
  \centering
  \begin{minipage}{0.65\textwidth}

  \begin{subfigure}[b]{0.49\textwidth}
    \centering
    \includegraphics[width=\textwidth]{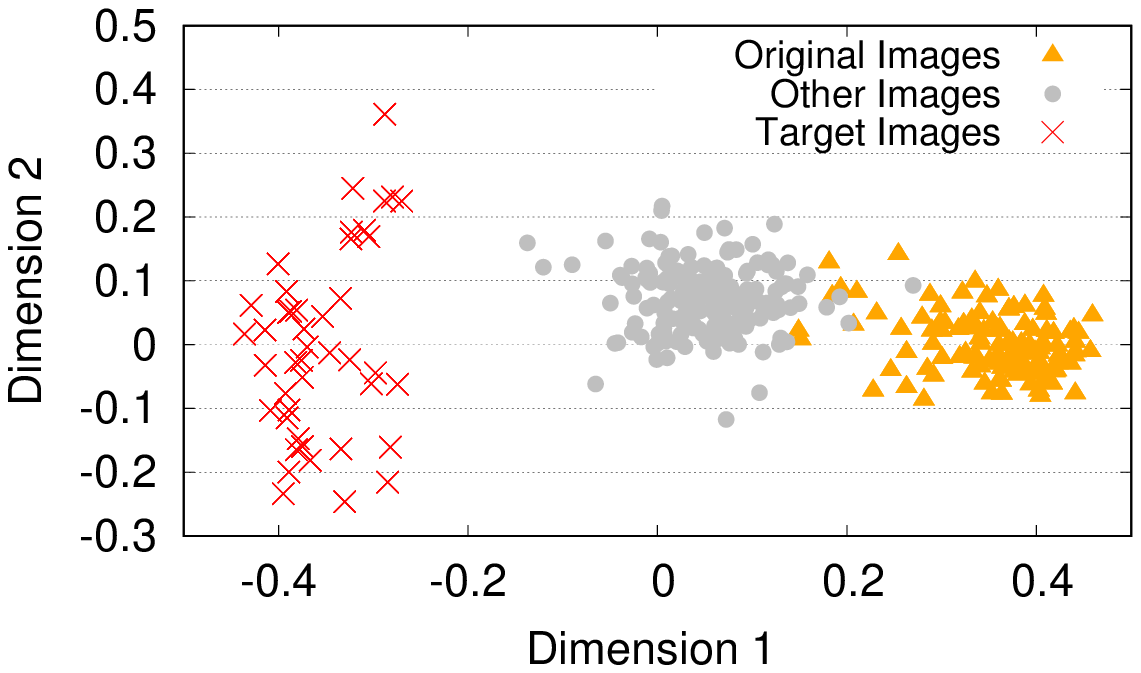}
    \caption{Before Cloaking}
  \end{subfigure}
  \centering
  \begin{subfigure}[b]{0.49\textwidth}
    \centering
    \includegraphics[width=\textwidth]{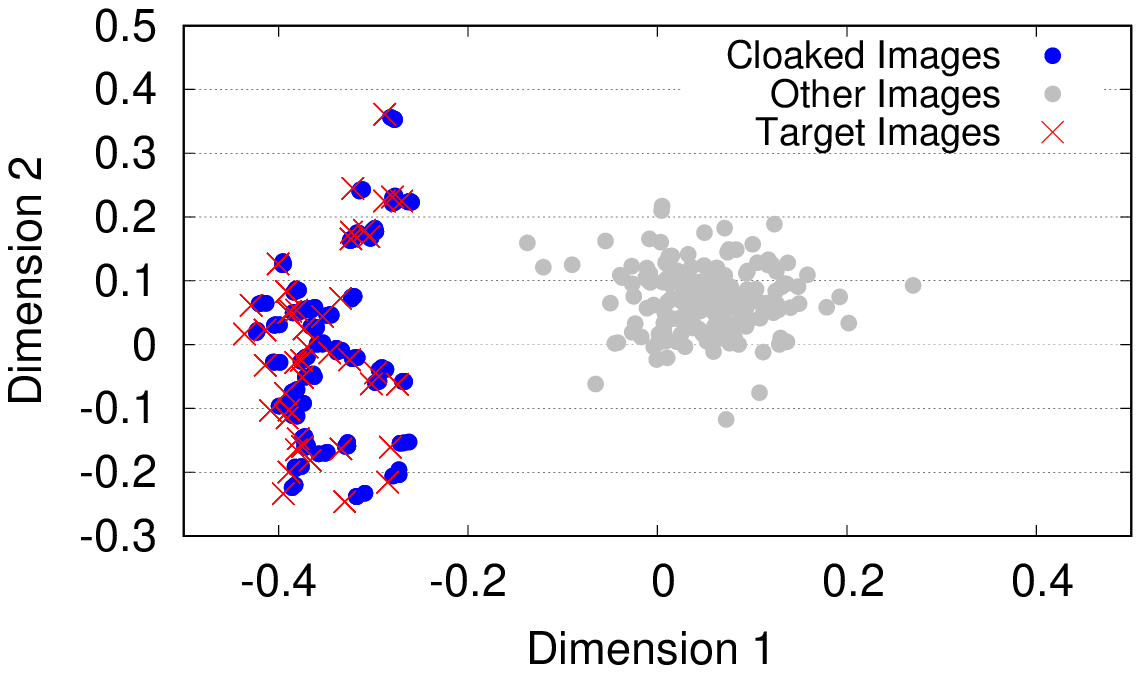}
    \caption{After Cloaking}
  \end{subfigure}
      \vspace{-0.1in}
  \caption{2-D PCA visualization of \vde{} feature space
    representations of user images (sampled from \facescrub{})
    before/after cloaking. Triangles are user's images, red
    crosses are target images, grey dots are images from another class. }
  \label{fig:pubfig_pca}
    \end{minipage}
    \hfill
    \begin{minipage}{0.32\textwidth}
    \centering
    \includegraphics[width=1\textwidth]{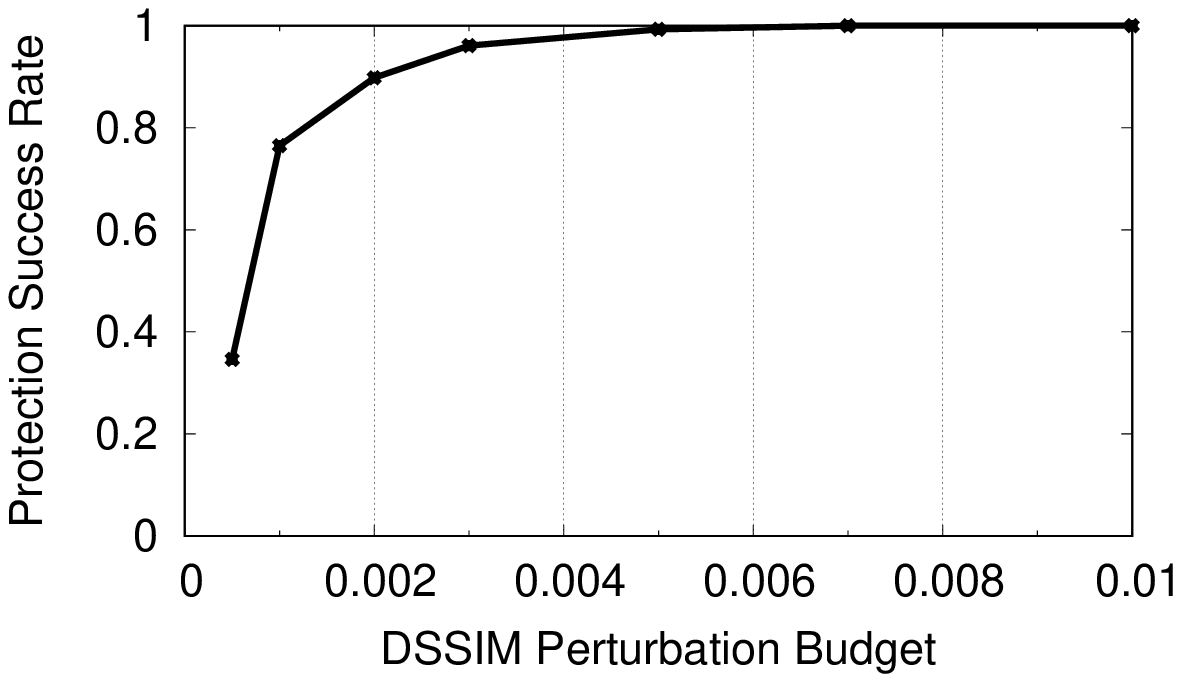}
    \vspace{0.003in}
    \caption{\emed{Protection performance as DSSIM perturbation budget
    increases. (User/Tracker: \win{})}}
    \label{fig:pubfig_perturbation}
  \end{minipage}
\end{figure*}

\secspace
\subsection{User/Tracker Sharing a Feature Extractor}
\label{sec:scenario1}

We start from the simple case where the user uses the same feature
extractor as the tracker to generate cloaks. 
We \emed{randomly select a label from \pubfig{} or \facescrub{}} to be the Fawkes user $U$. 
We then compute ``cloaks'' for a subset of $U$'s images, using each of the four feature
extractors in Table~\ref{tab:model_eval}.  On the tracker side, we perform
transfer learning on the {\em same} feature extractor (with cloaked images of
$U$) to build a model that recognizes $U$. Finally, we evaluate whether the
tracker model can correctly identify other {\em clean} images of $U$ it has
not seen before.

\begin{figure}[t]
      \centering
      \includegraphics[width=0.4\textwidth]{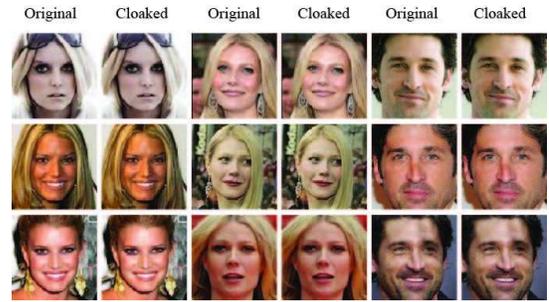}
      \caption{Pairs of original and cloaked images ($\rho = 0.007$). }
      \label{fig:sample_faces}
\end{figure}

Results show that  cloaking offers perfect protection, {\em i.e.} $U$
is always misclassified as
someone else, for all four feature extractors and under the perturbation budget $\rho$ = $0.007$.
To explore the impact of $\rho$,  Figure~\ref{fig:pubfig_perturbation} plots protection success rate
vs. $\rho$ when the tracker runs on the \facescrub{} dataset. Fawkes achieves $100\%$ protection success
rate when $\rho > 0.005$. Figure~\ref{fig:sample_faces} shows original
and cloaked images, demonstrating that cloaking does not
visually distort the original image. Even when $\rho = 0.007$, the perturbation is
barely detectable by the naked eye on a full size, color
image. For calibration, note that prior work~\cite{li2019hiding}
claims much higher DSSIM values (up to $0.2$) are imperceptible to the human eye.
Finally, the average $L2$ norm of our cloaks is $5.44$, which is smaller than that of
perturbations used in prior works~\cite{wang2018great,liu2016delving}.

\para{Feature Space Deviation.} The goal of a cloak is to change the image's
feature space representation in the tracker's model.  To examine the effect
of the cloak in the tracker model, we visualize feature space
representations of user images before and after cloaking, their
chosen target images, and \emed{a randomly chosen class from
the tracker's dataset}. We use principal
components analysis (PCA, a common dimensionality reduction technique)
to reduce the high dimensional feature space to 2
dimensions. Figure~\ref{fig:pubfig_pca} shows the PCA results for
cloaked images from a \pubfig{} class, using cloaks constructed on the \win{} feature
extractor. Figure~\ref{fig:pubfig_pca}(a) shows the feature space positions
of the original and target images before cloaking, along with a randomly
selected class. Figure~\ref{fig:pubfig_pca}(b) shows the updated feature
space after the original images have been cloaked. It is clear that feature
space representations of the cloaked images are well-aligned with those of
the target images, validating our intuition for cloaking
(an abstract view in Figure~\ref{fig:intuition}).

\para{Impact of Label Density.} As discussed in~\S\ref{sec:method},
the number of
labels present in the tracker's model impacts 
performance. When the tracker targets fewer labels, the feature space is
``sparser,'' and there is a greater chance the model continues to
associate the original feature space (along with the cloaked feature space)
with the user's label. We empirically evaluate the impact of fewer labels on
cloaking success using the \pubfig{} and \facescrub{} datasets ($65$ and
$530$ labels, respectively). We randomly sample $N$ labels (varying
$N$ from $2$ to $10$) to construct a model with fewer
labels. Figure~\ref{fig:pubfig_density} shows that for \pubfig{},
cloaking success rate grows from $68\%$ for $2$ labels to $>99\%$ for more
than $6$ labels, confirming that a higher label density improves
cloaking effectiveness.


\secspace
\subsection{User/Tracker Using Different Feature Extractors}
\label{sec:scenario2}

We now consider the scenario when the user and tracker use different 
feature extractors to perform their tasks. While the model transferability
property suggests that there are significant similarities in their respective
model feature spaces (since both are trained to recognize faces),
their  differences could still reduce the efficacy of
cloaking. Cloaks that shift image features significantly in one
feature extractor may produce a much smaller shift in a different
feature extractor.

\begin{figure*}[t]
  \centering
    \begin{minipage}{0.32\textwidth}
    \centering
    \includegraphics[width=1\textwidth]{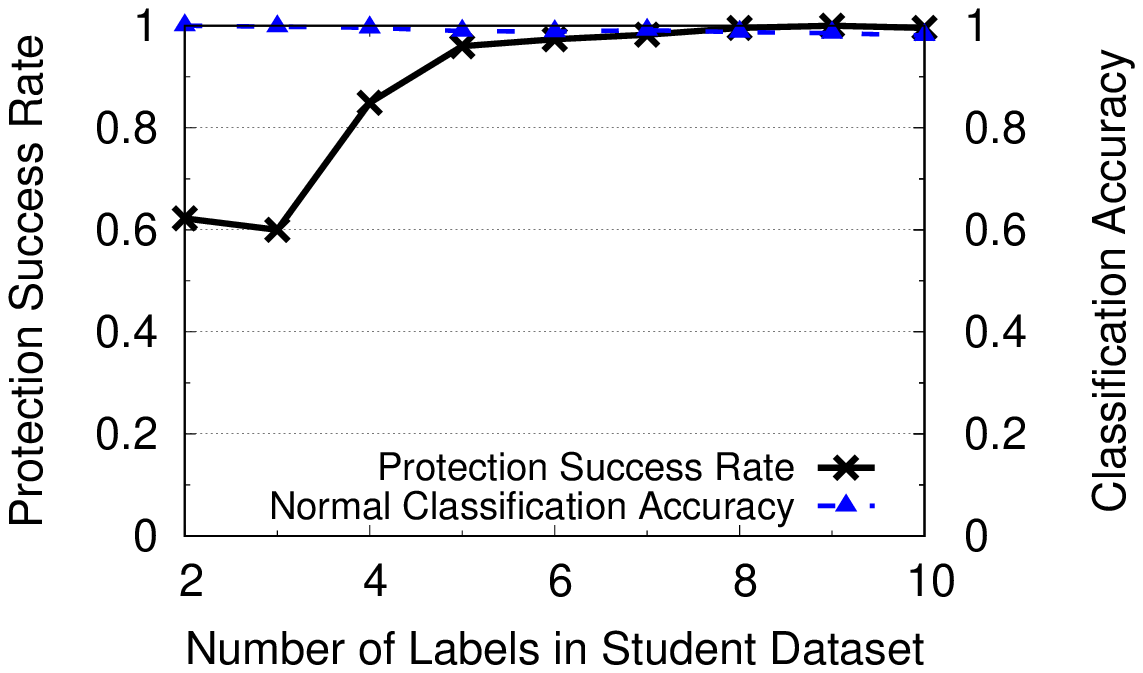}
    \vspace{-0.1in}
    \caption{Protection performance improves as the number of labels
    in tracker's model increases. (User/Tracker: \win{})}
    \label{fig:pubfig_density}
  \end{minipage}
  \hfill
  \begin{minipage}{0.32\textwidth}
    \centering
    \includegraphics[width=\textwidth]{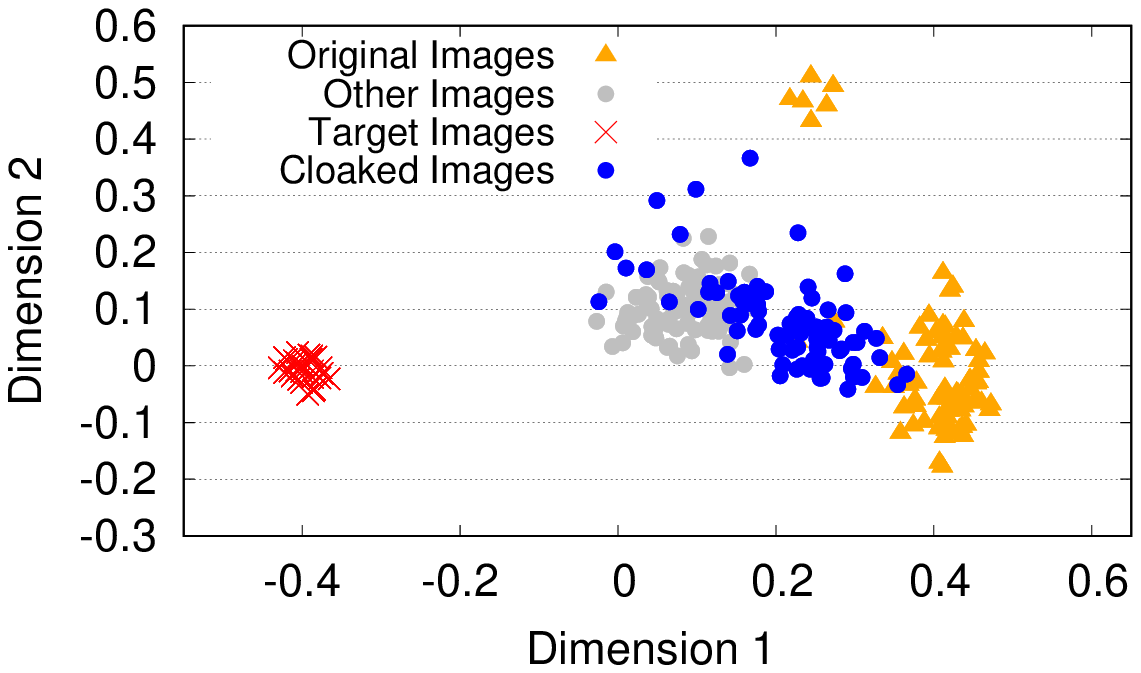}
    \vspace{-0.1in}
    \caption{\emed{Cloaking is less effective when users and trackers
    use different feature extractors. (User: \vde{}, Tracker: \win{})}}
  \label{fig:weak_pca}
  \end{minipage}
  \hfill
  \begin{minipage}{0.32\textwidth}
    \centering
    \includegraphics[width=\textwidth]{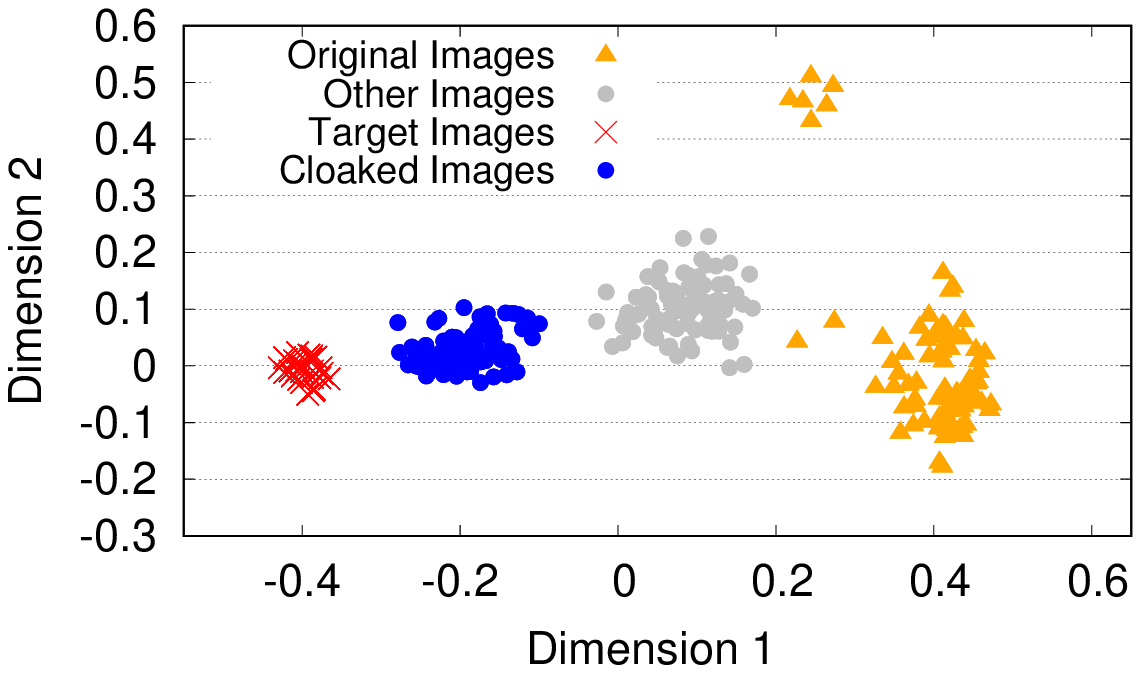}
    \vspace{-0.1in}
    \caption{\emed{Cloaks generated on robust
      models transfer better between feature
    extractors. (User: \vde{}, Tracker: \win{})}}
    \label{fig:robust_pca}
    \end{minipage}
\end{figure*}

To illustrate this, we empirically inspect
the change in feature representation between two different feature
extractors. We take the cloaked images (optimized using \vde{}), original images, and target
images from the \pubfig{} dataset and calculate their
feature representations in a {\em different} feature extractor,
\win{}. The result is visualized using two dimensional PCA and shown in
Figure~\ref{fig:weak_pca}. From the PCA visualization, the reduction
in cloak effectiveness is obvious. In the tracker's feature
extractor, the cloak ``moves'' the original image features only slightly
towards the target image features (compared to Figure~\ref{fig:pubfig_pca}(b)). 

\begin{table*}[t]
  \centering
  \resizebox{0.75\textwidth}{!}{
    \begin{tabular}{|l|c|c|c|c|c|c|c|c|}
    \hline
    \multirow{3}{*}{\textbf{\begin{tabular}[c]{@{}l@{}}User’s Robust \\ Feature \\ Extractor\end{tabular}}} & \multicolumn{8}{c|}{\textbf{Model Trainer's Feature Extractor}} \\ \cline{2-9} 
     & \multicolumn{2}{c|}{\vin{}} & \multicolumn{2}{c|}{{\vde{}}} & \multicolumn{2}{c|}{\win{}} & \multicolumn{2}{c|}{\wde{}} \\ \cline{2-9} 
     & PubFig & FaceScrub & PubFig & FaceScrub & PubFig & FaceScrub & PubFig & FaceScrub \\ \hline
    \vin{} & $100\%$ & $100\%$ & $100\%$ & $100\%$ & $95\%$ & $100\%$ & $100\%$ & $100\%$ \\ \hline
    \vde & $100\%$ & $100\%$ & $100\%$ & $100\%$ & $100\%$ & $100\%$ & $100\%$ & $100\%$ \\ \hline
    \win{}  & $100\%$ & $100\%$ & $100\%$ & $100\%$ & $100\%$ & $100\%$ & $99\%$ & $99\%$ \\ \hline
    \wde{} & $100\%$ & $100\%$ & $100\%$ & $100\%$ & $100\%$ & $97\%$ & $100\%$ & $96\%$ \\ \hline
    \end{tabular}
  }
  \vspace{-0.05in}
   \caption{\emed{Protection performance of cloaks generated on robust
       feature extractors}.}
  \label{tab:robust_teacher}
\end{table*}

\para{Robust Feature Extractors Boost Transferability.} To address the
problem of cloak transferability, we draw on recent work linking model
robustness and transferability. Demontis \etal~\cite{demontis2019adversarial}
argue that an input perturbation's (in our case, cloak's) ability to transfer between
models depends on the ``robustness'' of the feature extractor used to
create it. \emed{They show that more ``robust'' models are less
reactive to small perturbations on inputs. Furthermore, they claim
that perturbations (or, again, cloaks) generated on more robust models will take on ``universal'' characteristics that are able to effectively fool other models.}


Following this intuition,  we propose to improve cloak transferability
by increasing the user
feature extractor's robustness. 
This is done by  applying {\em adversarial
  training}~\cite{madry2017towards,goodfellow2014explaining}, which trains the model on perturbed data to make it less
  sensitive to similar small perturbations on inputs.
Specifically, for each feature extractor,  we generate adversarial
examples using the PGD attack~\cite{kurakin2016adversarial}, a widely
used method for adversarial training. \shawnedit{Following prior 
work~\cite{madry2017towards}, we run the PGD\footnote{We found that robust models trained on CW
attack samples~\cite{carlini} produce similar results} algorithm for $100$ steps
using a step size of $0.01$.}
We train each feature extractor for an additional $10$
epochs. These updated feature extractors are then used to generate user cloaks on the \pubfig{} and \facescrub{} datasets. 

Results in  Table~\ref{tab:robust_teacher}  show that each robust feature extractor produces cloaks that transfer almost perfectly to
the tracker's models.   Cloaks now have protection success rates $>95\%$ when the tracker
uses a different feature extractor. We visualize their feature representation using
PCA in Figure~\ref{fig:robust_pca} and see that, indeed, cloaks
generated on robust extractors transfer better than cloaks computed on
normal ones.

\secspace
\subsection{Tracker Models Trained from Scratch}
\label{sec:scenario3}

Finally, we consider the scenario in which a
powerful tracker trains their model from scratch. We select the user
$U$ to be a label inside the
\webface{} dataset. We generate cloaks on user images using \shawnedit{the robust \vin{} feature
extractor from \S\ref{sec:scenario2}.}  The tracker 
then uses the \webface{} dataset (but $U$'s cloaked images) to train
their model from scratch. Again our cloas achieve a success rate of $100\%$.
Other combinations of labels and user-side feature generators all
have $100\%$ protection success. 

\secspace
\section{Image Cloaking in the Wild}
\label{sec:real}


Our results thus far have focused on limited configurations, including
publicly available datasets and known model architectures.
Now, we wish to understand the performance of Fawkes on deployed facial recognition
systems in the wild. 

We evaluate the real-world effectiveness of image cloaking by applying Fawkes
to photos of one of the co-authors. We then intentionally leak a portion of
these cloaked photos to public cloud-based services that perform
facial recognition,
including Microsoft Azure Face~\cite{microsoft_api}, Amazon
Rekognition~\cite{aws_api}, and Face++~\cite{face_plus_plus}. These are the
global leaders in \emed{facial recognition and their
  services} are used by businesses, police, private entities, and governments in the US and Asia.

\secspace
\subsection{Experimental Setup}
We manually collected $82$ high-quality pictures of a co-author that feature
a wide range of lighting conditions, poses, and facial expressions. We
separate the images into two subsets, one set of 50 images for ``training''
and one set of 32 images for
``testing.'' We generate both normal and robust cloaks for the
``training'' images using the setup discussed in
Section~\ref{sec:eval} (using normal and robust versions of the \win{} feature extractor). This allows us
to compare the relative effectiveness of normal and robust user feature
extractors in real life.

For each API service, we experiment with  three scenarios:
\begin{packed_itemize} \vspace{-0.05in}
\item {\bf Unprotected:} We upload original training images, and 
  test the model's classification accuracy on testing images.
\item {\bf Normal Cloak:} We upload training images protected by a
  {\em nonrobust} cloak and then test the model's
  classification accuracy on the testing images.
\item {\bf Robust Cloak:} We upload training images protected by a
  {\em robust} cloak and test the model's classification accuracy on the testing images.
\end{packed_itemize}\vspace{-0.05in}

For each scenario, we use the online service APIs to upload training images to
the API database, and then query the APIs using the uncloaked testing images.
The reported protection success rate is the proportion of uncloaked test
images that the API fails to correctly identify as our co-author.


\begin{table}[t]
  \centering
  \resizebox{\columnwidth}{!}{
    \begin{tabular}{|l|c|c|c|}
    \hline
    \multirow{2}{*}{\textbf{\begin{tabular}[c]{@{}l@{}}Face \\Recognition\\ API\end{tabular}}} & \multicolumn{3}{c|}{\textbf{Protection Success Rate}} \\ \cline{2-4} 
     & \textbf{\begin{tabular}[c]{@{}c@{}}Without \\ protection\end{tabular}} & \textbf{\begin{tabular}[c]{@{}c@{}}Protected by \\ normal cloak\end{tabular}} & \textbf{\begin{tabular}[c]{@{}c@{}}Protected by \\ robust cloak\end{tabular}} \\ \hline
    \begin{tabular}[c]{@{}l@{}}Microsoft Azure \\ Face API\end{tabular} & $0\%$ & $100\%$ & $100\%$ \\ \hline
    \begin{tabular}[c]{@{}l@{}}Amazon Rekognition \\ Face
      Verification\end{tabular} & $0\%$ & $34\%$ & $100\%$ \\ \hline
    \begin{tabular}[c]{@{}l@{}}Face++ \\ Face Search API\end{tabular} & $0\%$ &
                                                                              $0\%$ & $100\%$ \\ \hline
    \end{tabular}
  }
  \caption{\emed{Cloaking is highly effective against cloud-based face
    recognition APIs (Microsoft, Amazon and Face++). }}
  \label{tab:real_world_eval}
  \vspace{-0.1in}
\end{table}

\secspace
\subsection{Real World Protection Performance}

\para{Microsoft Azure Face API.} Microsoft Azure Face
API~\cite{microsoft_api} is part of Microsoft Cognitive Services, and is
reportedly used by many large corporations including Uber and Jet.com. The
API provides face recognition services. A client uploads training images of
faces, and Microsoft trains a model to recognize these faces. The API has a
``training'' endpoint that must be called before the model will recognize
faces, which leads us to believe that Microsoft uses transfer learning to
train a model on user-submitted images.

Our normal cloaking method is 100\% effective against the Microsoft Azure
Face API. Our robust cloaks also provide 100\% protection against the Azure
Face API.  Detailed protection results are shown in
Table~\ref{tab:real_world_eval}.


\para{Amazon Rekognition Face Verification. } Amazon Rekognition~\cite{aws_api}
provides facial search services that the client can use to detect,
analyze, and compare faces. The API is used by various large
corporations including the NFL, CBS, and National Geographic, as well as
law enforcement agencies in Florida and Oregon, and the U.S. Immigration and
Customs Enforcement agency (ICE). 

It is important to note that Amazon Rekognition does not specifically
train a neural network to classify queried images. Instead, it computes an image
similarity score between the queried image and the ground truth images
for all labels. If the similarity score exceeds a threshold for some label, Amazon
returns a match. Our cloaking technique is not designed to fool a tracker who
uses similarity matching. However, we believe our cloaking technique
should still be effective against Amazon Rekognition, since
cloaks create a feature space separation between original and cloaked
images that should result in low similarity scores between them. 


Table~\ref{tab:real_world_eval} shows that our normal cloaks only achieve
a protection success rate of $34\%$.  However, our robust cloaks
again achieve a $100\%$ protection success rate.


\para{Face++. } Face++~\cite{face_plus_plus} is a well-known face recognition
system developed in China that
claims to be extremely robust against a variety of attacks
({\em i.e.} adversarial masks, makeup, etc.). Due to its high performance
and perceived robustness, Face++ is widely used by financial services
providers and other security-sensitive customers. Notably, Alipay uses
Face++'s services to authenticate users before processing
payments. Lenovo also uses Face++ services to perform face-based
authentication for laptop users. 

Our results show that normal cloaking is completely ineffective against
Face++ ($0\%$ protection success rate; see
Table~\ref{tab:real_world_eval}). This indicates that their model is indeed
extremely robust against input perturbations. \emed{However, as
  before, our robust cloaks achieve a $100\%$ success rate.}



\para{Summary.} Microsoft Azure Face API, Amazon Rekognition and
Face++ represent three of the most popular and widely deployed
facial recognition services today. The success of Fawkes cloaking
techniques suggests our approach is realistic and practical against
production systems. While we expect these systems to continue
improving, we expect cloaking techniques to similarly evolve over time to keep pace.

\secspace
\section{Trackers with Uncloaked Image Access}
\label{sec:sybil}

Thus far we have assumed that the tracker only has access to {\em cloaked
  images} of a user, {\em i.e.} the user is perfect in applying her cloaking
protection to her image content, and disassociating her identity from images
posted online by friends. In real life, however, this may be too strong an
assumption. Users make mistakes, and unauthorized labeled images of the user
can be taken and published online by third parties such as newspapers and
websites.

In this section, we consider the possibility of the tracker obtaining leaked,
uncloaked images of a target user, {\em e.g.} Alice. We first evaluate the
impact of adding these images to the tracker's model training data. We then
consider possible mechanisms to mitigate this impact by leveraging the use
of limited sybil identities online.

\secspace
\subsection{Impact of Uncloaked Images}
Intuitively, a tracker with access to some labeled, uncloaked images of a
user has a much greater chance of training a model $M$
that successfully recognizes clean images of \dave{that user}. Training a model with
both cloaked and uncloaked user images means the model will observe a
much larger spread of features all designated as \dave{the user}. Depending on how $M$
is trained and the presence/density of other labels, it can a) classify both
regions of features as \dave{the user}; b) classify both regions and the region
between them as \dave{the user}; or c) ignore these feature dimensions and identify
\dave{the user} using some alternative features ({\em e.g.} other facial features) that
connect both uncloaked and cloaked versions of \dave{the user's images}.

\begin{figure*}[t]
  \centering
  \begin{minipage}{0.32\textwidth}
    \centering
    \includegraphics[width=1\textwidth]{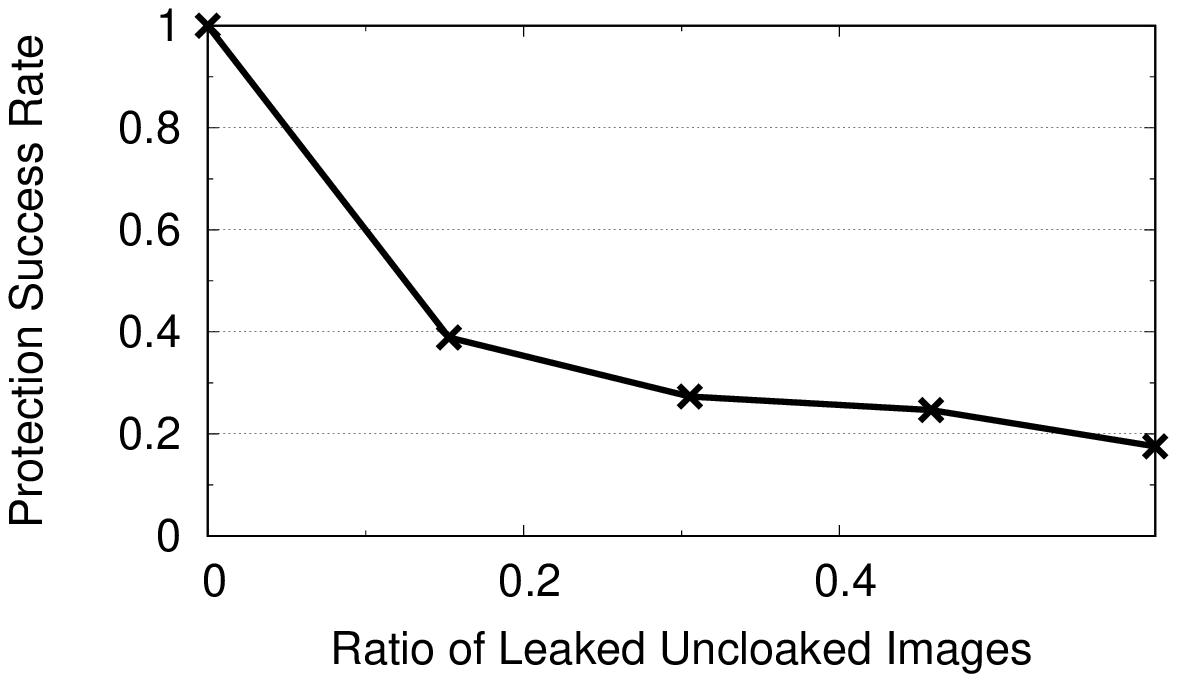}
    \vspace{-0.1in}
    \caption{\emed{Protection success rate decreases when the tracker
        has more original user images. (User/Tracker: \win{})}}
    \label{fig:sybil_base}
  \end{minipage}
  \hfill
  \begin{minipage}{0.32\textwidth}
    \centering
    \includegraphics[width=1\textwidth]{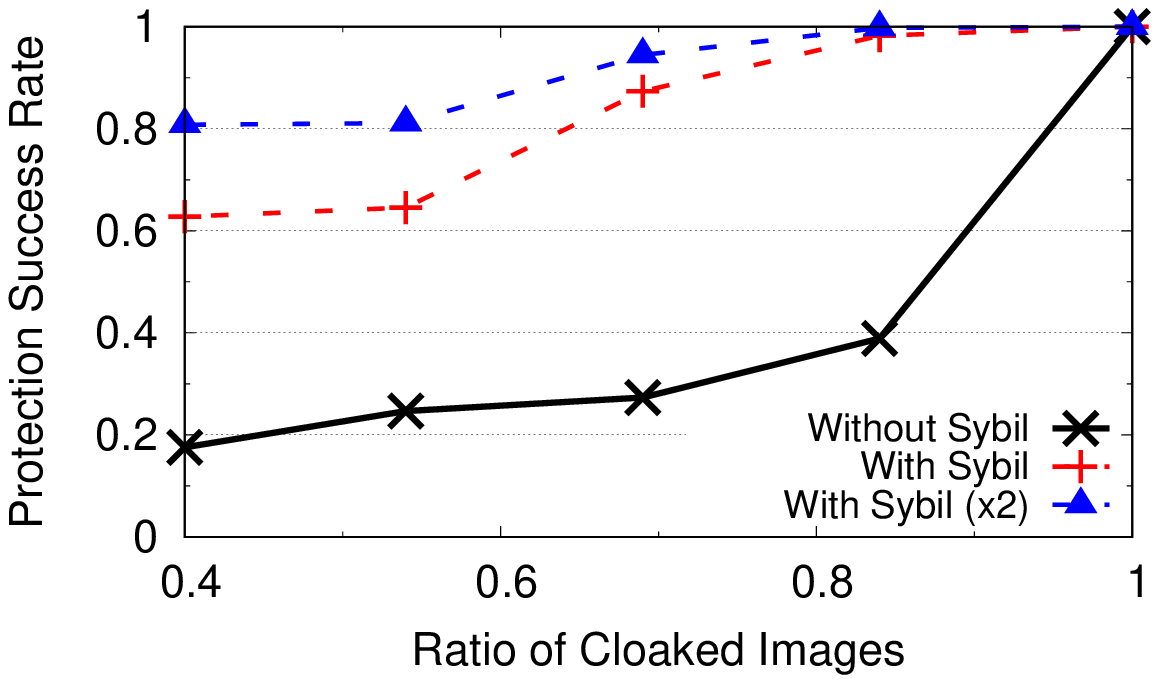}
    \vspace{-0.1in}
    \caption{\emed{Protection success rate is high when the user has a Sybil
      account, even if tracker has original user
      images. (User/Tracker: \win{})}}
    \label{fig:scrub_sybil}
  \end{minipage}
  \hfill
  \begin{minipage}{0.32\textwidth}
    \centering
    \includegraphics[width=1\textwidth]{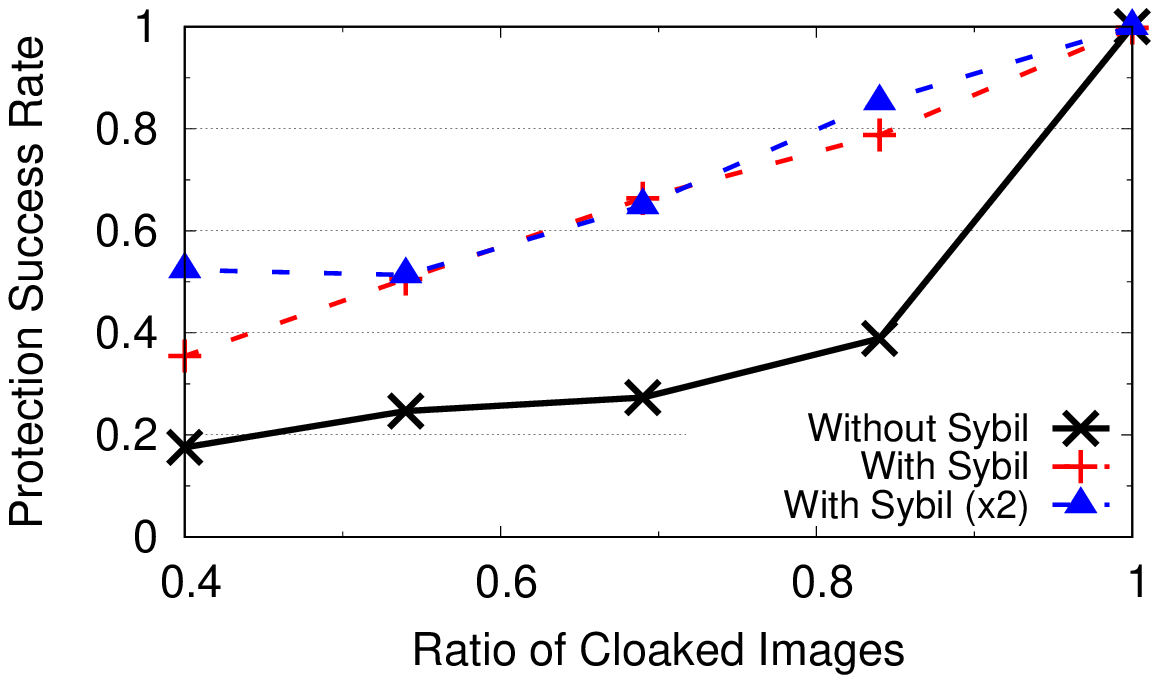}
    \vspace{-0.1in}
    \caption{\emed{Sybils jointly optimized on
      four feature extractors have reasonably high protection
      success for each individual extractor.}}
    \label{fig:jointly}
  \end{minipage}
\end{figure*}

We assume the tracker cannot visually distinguish between cloaked
and uncloaked images and trains their model on both.
We quantify the impact of training with uncloaked images using a simple test with
cloaks generated from~\S\ref{sec:scenario1} and a model trained on both
cloaked and uncloaked images. Figure~\ref{fig:sybil_base} shows the drop
in protection success for \facescrub{} dataset
as the ratio of uncloaked images in the training dataset increases. 
The protection success rate drops below $39\%$ when more than
$15\%$ of the user's images are uncloaked.

Next, we consider proactive mitigation strategies against leaked images. The
most direct solution is to intentionally release more cloaked images,
effectively flooding a potential tracker's training set with cloaked images
to dominate any leaked uncloaked images. In addition, we consider the use of
a cooperating secondary identity (more details below). For
simplicity, we assume that: trackers have access to a {\em small} number of a
user's uncloaked images; the user is unaware of the contents of the uncloaked
images obtained by the tracker; and users know the feature extractor used by
the tracker.

\secspace
\subsection{Sybil Accounts}

In addition to proactive flooding of cloaked images, we explore the use of 
cooperative {\em Sybil accounts} to induce model misclassification. A Sybil
account is a separate account controlled by the user that exists in the same
Internet community (\ie Facebook, Flickr) as the original account. Sybils
already exist in numerous online communities~\cite{yang2014uncovering}, and
are often used by real users to curate and compartmentalize content for
different audiences~\cite{sybil-normal}.  While there are numerous detection
techniques for Sybil detection, individual Sybil accounts are difficult to
identify or remove~\cite{wang2013you}.

In our case, we propose that privacy-conscious users create a secondary
identity, preferably not connected to their main identity in the metadata or
access patterns. Its content can be extracted from public sources, from a
friend, or even generated artificially via generative adversarial networks
(GANs)~\cite{gans-nyt}. Fawkes modifies Sybil images (in a manner similar to
cloaking) to provide additional protection for the user's original
images. Since \emed{Sybil and user images} reside in the same
communities, we expect trackers will collect \emed{both}. While there
are powerful re-identification techniques that could be used to associate the Sybil back to the original user, we assume they are
impractical for the tracker to apply at scale to its population of tracked
users. 

\begin{figure}[t]
  \centering
  \includegraphics[width=1\columnwidth]{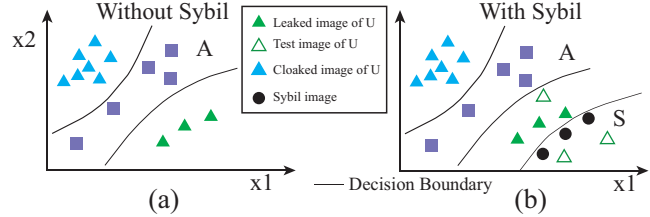}
  \vspace{-0.15in}
  \caption{Intuition behind Sybil integration visualized in a 2D
    feature space. Without Sybils, a tracker's model will use
    leaked training images of $U$ to learn $U$'s true feature space (left),
    leading to the correct classification of images of $U$. Sybil images $S$
    complicate the model's decision boundary and cause misclassification of
    $U$'s images, even when leaked images of $U$ are present (right). }

  \label{fig:sybil_illu}
  \vspace{-0.1in}
\end{figure}

\para{Sybil Intuition.} To bolster cloaking effectiveness, the user modifies
Sybil images so they occupy the same feature space as a user's uncloaked
images. These Sybil images help confuse a model trained on both Sybil images
{\em and} uncloaked/cloaked images of a user, increasing the protection
success rate.  Figure~\ref{fig:sybil_illu} shows the high level
intuition. Without Sybil images, models trained on a small portion of
uncloaked (leaked) images would easily associate test images of the user with
the user's true label (shown on left). Because the leaked uncloaked images
and Sybil images are close by in their feature space representations, but
labeled differently ({\em i.e.} `` User 1'' and ``User 2''), the tracker
model must create additional decision boundaries in the feature space (right
figure). These additional decision boundaries decrease the likelihood of
associating the user with her original feature space.

For simplicity, we explore the base case where the user is able to obtain one
single Sybil identity to perform feature space obfuscation on her behalf. Our
technique becomes even more effective with multiple Sybils, but provides much
of its benefit with images labeled with a single Sybil identity.

\para{Creating Sybil images. } Sybil images are created by adding a specially
designed cloak to a set of candidate images. Let $x_C$ be an
image from the set of candidates the user obtains (\ie images
generated by a GAN) to populate the Sybil account. To create the final
Sybil image, we create a cloak $\delta(x_C, x)$ that minimizes the
feature space separation between $x_C$ and user's original image $x$, for each candidate. The optimization is equivalent to setting $x$
as the target and optimizing to create $x_C \oplus \delta(x_C,x)$ as
discussed in~\S\ref{sec:design}.  After choosing the final $x_c$ from
all the candidates,  a ready-to-upload Sybil image $x_S=x_C \oplus
\delta(x_C, x)$.


\secspace
\subsection{Efficacy of Sybil Images}
Sybil accounts can increase a user's protection success rate when the tracker
controls a small number of a user's uncloaked images. To experimentally
validate this claim, we choose a label from the tracker's dataset to be the
Sybil account (controlled by the user), and split the user's images into two
disjoint sets: $A$ contains images that were processed by Fawkes, and whose
cloaked versions have been shared online; and $B$ contains original images
leaked to the tracker. For each synthetic image of the Sybil, we randomly
select an uncloaked image of the user in set $A$. \shawnedit{We select
  one Sybil image per uncloaked image in $A$.} Then, we cloak all the
candidate images using the methodology discussed in~\S\ref{sec:design}. The
resulting Sybil images mimic the feature space representation of uncloaked
user images. From the tracker's perspective, they have access to cloaked user
images from set $A$, uncloaked images from set $B$, and the Sybil images.

Figure~\ref{fig:scrub_sybil} compares the protection success rate with and
without Sybil accounts (with \win{} as user's and tracker's feature
extractor). The use of a Sybil account significantly improves the protection
success rate when an attacker has a small number of original images. The
protection success rate remains above $87\%$ when the ratio of the original
images owned by the tracker is less than $31\%$. 

As discussed, a user can create as many Sybil images as they
desire. When the user uploads more Sybil images, the protection success rate
increases. Figure~\ref{fig:scrub_sybil} shows that when the
user has uploaded \shawnedit{$2$ Sybil images per uncloaked image}, the protection success rate increases by $5.5\%$.

\shawnedit{
\para{Jointly Optimize Multiple Feature Extractors. } The user may not
know the \emed{tracker's exact feature extractor. However,} given the
small number of face feature extractors \emed{available online}, she
is likely to know that the tracker would use one of several candidate
feature extractors. Thus, she could jointly optimize the Sybil cloaks
\emed{to simultaneously fool all the candidate feature extractors}.

We test this in a simple experiment by jointly optimizing Sybil cloaks
on the four feature extractors from
~\S\ref{sec:eval}. We evaluate the cloak's performance when the
tracker uses one of the four. Figure~\ref{fig:jointly} shows the Sybil effectiveness averaged across the $4$ feature extractors. The average protection success rate remains above $65\%$ when the ratio of the original images owned by the tracker is less than $31\%$.}

\secspace
\section{Countermeasures}
\label{sec:counter}

\begin{figure*}[t]
  \centering
  \begin{minipage}{0.32\textwidth}
    \centering
    \includegraphics[width=1\textwidth]{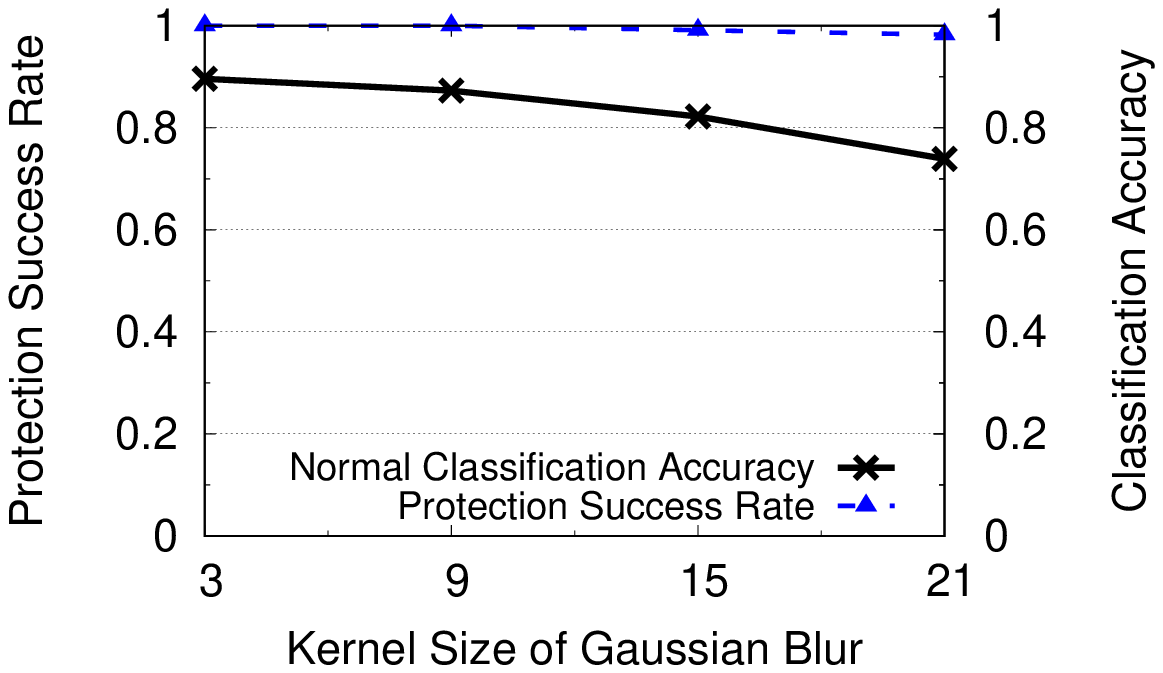}
    \vspace{-0.1in}
    \caption{\emed{Normal
        classification accuracy decreases as input blurring increases
  but protection success rate remains high.}}
    \label{fig:blurring}
  \end{minipage}
  \hfill
  \begin{minipage}{0.32\textwidth}
    \centering
    \includegraphics[width=1\textwidth]{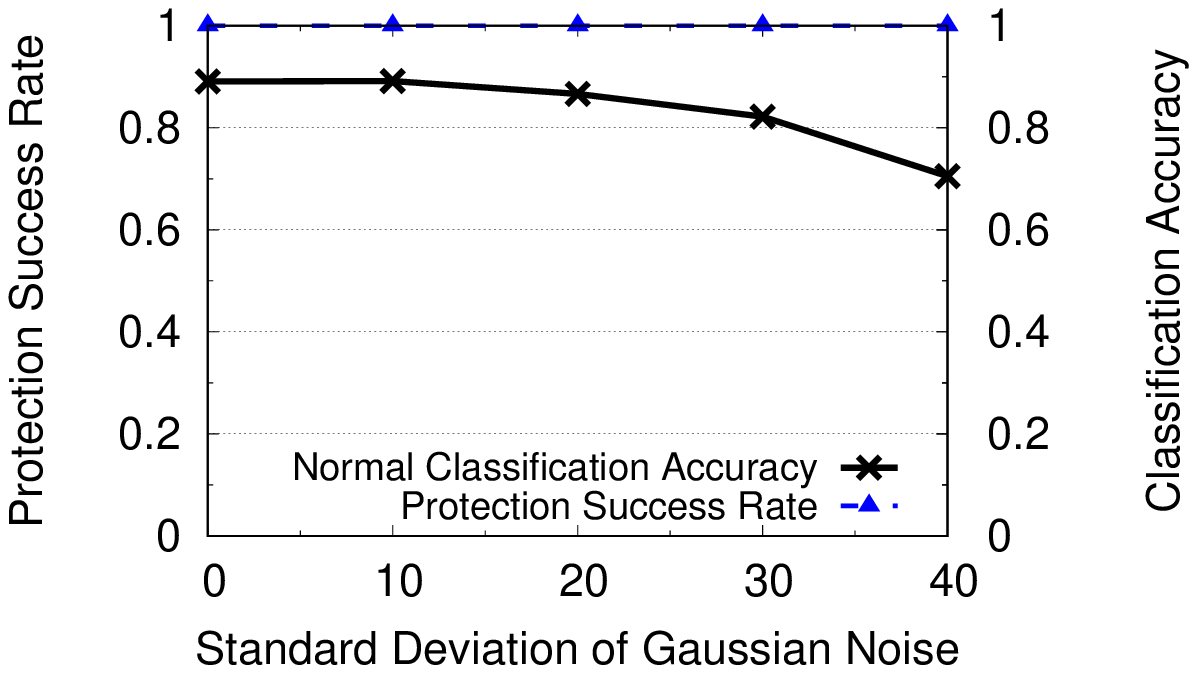}
    \vspace{-0.1in}
    \caption{\emed{Normal classification
      accuracy decreases as Gaussian noise is added to
      inputs but protection success rate remains high.}}
    \label{fig:noise}
  \end{minipage}
  \hfill
  \begin{minipage}{0.32\textwidth}
    \centering
    \includegraphics[width=1\textwidth]{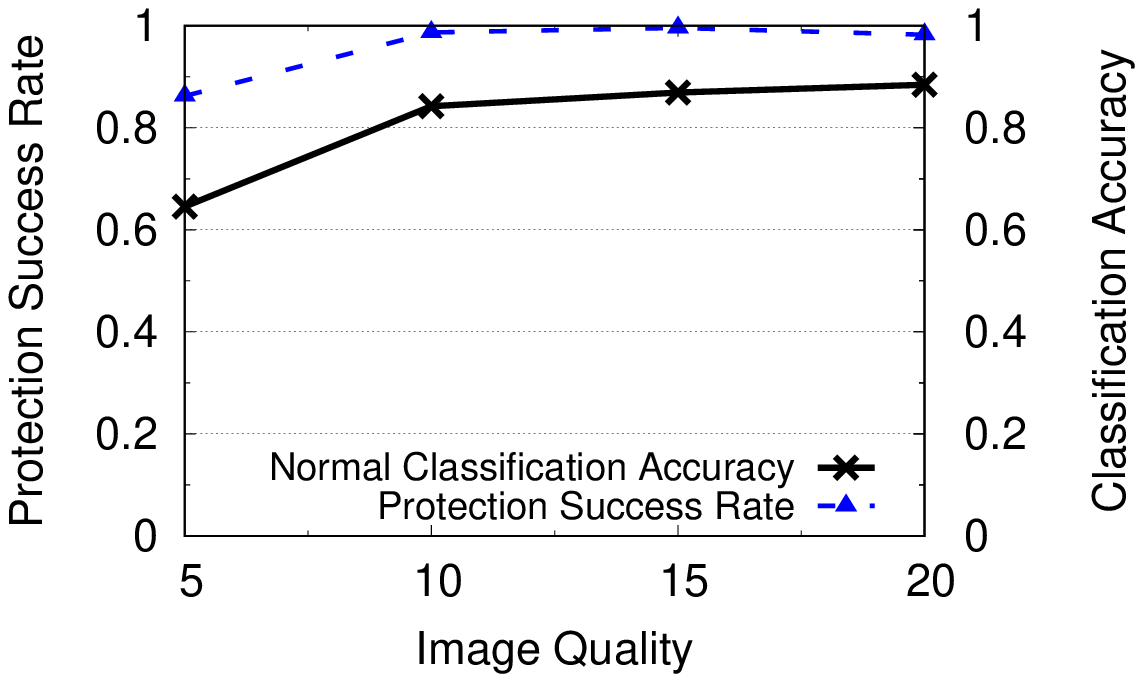}
    \vspace{-0.1in}
    \caption{\emed{Protection success rate and normal classification
      accuracy increase as image quality increases using JPEG compression.}}
    \label{fig:compress}
  \end{minipage}
  \hfill
\end{figure*}

In this section, we explore potential countermeasures a tracker could
\emed{employ to reduce the effectiveness of} image cloaking. We consider and (where
possible) empirically validate methods to {\em remove} cloaks from images, as
well as techniques to detect the presence of cloak perturbations on images.
Our experiments make the strongest possible assumption about the tracker:
that they know the precise feature extractor a user used to optimize cloaks.
We test our countermeasures on a tracker's model trained on the \facescrub{}
dataset. Cloaks were generated using the same robust \vde{} feature
extractor from~\S\ref{sec:scenario2}.

\para{Inherent Limits on Cloaking Success. } We acknowledge that cloaking
becomes less effective when an individual is an {\em active target} of a
tracker. If a tracker strongly desires to train a model that recognizes a
certain individual, they can take drastic measures that cloaking cannot
withstand. For example, a tracker could learn their movements or invade their
privacy ({\em i.e.} learn where they live) by following them physically.

\secspace
\subsection{Cloak Disruption}

\shawnedit{Without knowing which images in the dataset are cloaked,
the tracker may utilize the following techniques to disrupt Fawkes'
protection performance, 1) transforming images or 2) deploying an extremely robust model. We present and evaluate Fawkes's performance against these two potential countermeasures. }

\para{Image Transformation.} A simple technique to mitigate the impact
of small image perturbations is to transform images in the training dataset before
using them for model training~\cite{carlini2017adversarial,feinman2017detecting}. These transformations include image
augmentation, blurring, or adding noise. Additionally, images posted
online are frequently compressed before sharing (i.e. in the
upload process), \emed{which could impact cloak efficacy.}

\emed{However, we find that \shawnedit{none} of these transformations defeat our cloaks.} The protection
success rate remains $100\%$ even when data augmentation is applied
to cloaked images~\footnote{Image augmentation   parameters: rotation
  range=$20^o$, horizontal shift=$15\%$, vertical   shift=$15\%$, zoom
  range=15\%}. \emed{Applying Gaussian
blurring degrades normal accuracy by up to $18\%$ (as kernel size
increases) while cloak protection success rate remains $ > 98\%$ (see
Figure~\ref{fig:blurring}).} \emed{Adding Gaussian noise to
  images merely disrupts normal classification accuracy -- the cloak protection success rate remains
  above $100\%$ as the standard deviation of the noise distribution
  increases (see Figure~\ref{fig:noise})}. 
\emed{Even image compression cannot defeat our
    cloak}. We use progressive JPEG~\cite{wallace1992jpeg}, reportedly
  used by Facebook and Twitter, to
  compress the images in our dataset. The image quality, as standard by Independent JPEG Group~\cite{jpeg_standard}, ranges from $5$ to $95$ \emed{(lower value = higher compression)}. As shown in Figure~\ref{fig:compress}, image compression decreases the
protection success rate, but more significantly degrades normal
classification accuracy.

\begin{figure}[t]
    \centering
    \includegraphics[width=0.4\textwidth]{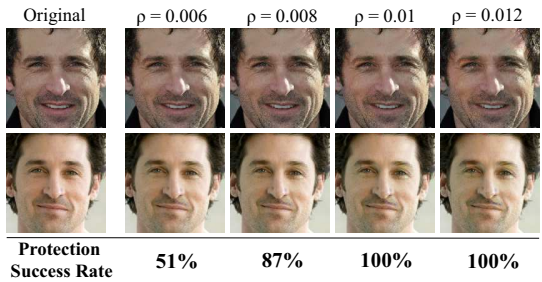}
    \vspace{-0.1in}
    \caption{\emed{When the user's feature extractor is much less robust
      than the tracker's feature extractor, the user can improve their
      protection success rate by increasing their DSSIM
      budget. (User: \vde{}, Tracker: \win{})}}
    \label{fig:perturbation_robust}
      \vspace{-0.1in}
\end{figure}


\para{Robust Model. } As shown in~\S\ref{sec:eval}, cloaks
constructed on robust feature extractors transfer well to trackers'
less robust feature extractors. Thus, a natural countermeasure a
tracker could employ is training \emed{their model to be extremely robust.}



Despite the \emed{theoretically proven trade-off} between normal accuracy and robustness~\cite{tsipras2018robustness}, future work may find a
way to improve model robustness while minimizing the accompanying drop
in accuracy. Thus, we evaluate cloaking success when the
tracker's model is much more robust than the user's
feature extractor. In our simplified test, the user has \emed{a}
robust \vde{} feature extractor (adversarially
trained for \textbf{$3$} epochs), \emed{while} the tracker has an
extremely robust \win{} feature extractor (adversarially trained for
\textbf{$20$} epochs). \emed{When the tracker's model is this robust}, the user's cloak only achieves a $64\%$ protection success rate.

However, if the user is extremely privacy sensitive, she could
increase the visibility of her cloak perturbation to achieve a higher
protection success rate. Figure~\ref{fig:perturbation_robust}
\emed{highlights the} trade off between protection \emed{success} and the input DSSIM
level. The cloak's protection success rate increases to $100\%$
once the DSSIM perturbation is $> 0.01$. 

\secspace
\subsection{Cloak Detection}

We now propose techniques a tracker could employ to detect cloaked images in
their dataset. We also discuss mitigations the user could apply to avoid
detection. 

\para{Existing Poison Attack Detection.} \emed{Since cloaking is a form of data poisoning, prior
work on detecting poisoning
attacks}~\cite{gupta2019strong,steinhardt2017certified,paudice2018detection,wang2019neural,chen2018detecting,shen2016auror}
could be helpful. However, all prior works assume that
poisoning only affects a small percentage of training images, making outlier
detection useful. Fawkes poisons an
entire model class, rendering outlier detection useless by removing
the correct baseline. 

\para{Anomaly Detection w/o Original Images. } We first consider
anomaly detection techniques in the scenario where the tracker does
not have \emed{any original user} images. If
trackers obtain both target and cloaked user images, they can
detect \emed{unusual closeness between cloaked images and target images in
model feature space}. Empirically, the $L2$ feature space distance
between the cloaked class centroid and the target class centroid is
$3$ standard deviations \emed{smaller than the mean separation of other classes}. Thus, user's cloaked images can be detected. 

However, a user can trivially overcome this detection by
maintaining separation between cloaked and target images
during cloak optimization. \emed{To show this, we use} the same
experimental setup as in~\S\ref{sec:scenario1} but terminate the
cloak optimization once a cloaked image is $20\%$ of the original $L2$
distance from the target image. The cloak still achieves a
$100\%$ protection success rate, but the \emed{cloak/target separation
  remains large enough to evade the previous detection method.}




\para{Anomaly Detection w/ Original Images. }
When the trackers have access to original training images
(see~\S\ref{sec:sybil}), they could~\emed{use clustering to
  see if there are two distinct feature clusters associated with the user's images (i.e. cloaked and uncloaked). Normal classes should have only
  one feature cluster.} To \emed{do this, the tracker could run a 2-means clustering on each class's feature space, flagging
  classes with two distinct centroids as potentially cloaked. When we
  run this experiment, we find that the distance between the two centroids of 
  a protected user class is $3$ standard deviations larger than the average
  centroid separation in normal classes. In this way, the tracker
  can use original images to detect the presence of cloaked images.}



To reduce the probability of detection by this method, the user can choose a target
class that does not create such a large feature space separation. We
empirically evaluate this mitigation strategy using the same
experimental configuration as in~\S\ref{sec:scenario1} but choose a target label with
average (rather than maximal) distance from their class. The cloak
generated with this method still achieves a $100\%$ protection success
rate, but $L2$ distance between the two cluster centroids is
within $1$ standard deviation of average. 

\revise{\dave{The user can evade this anomaly detection strategy using the maximum distance optimization}
  strategy in \S\ref{sec:design}. In practice, for any tracker model with a
  moderate number of labels (>$30$), cloaks generated with average or maximum
  difference optimization consistently achieves high cloaking success. Our
  experimental results show these two methods perform identically in
  protection success against both our local models and the Face++ API.  
}

\secspace
\section{Discussion and Conclusion}
\label{sec:discussion}
\vspace{-0.06in}

In this paper, we present a first proposal to protect
\dave{individuals} from \dave{recognition by} unauthorized
and unaccountable facial recognition systems. Our approach applies small,
carefully computed perturbations to cloak images, so that they are shifted
substantially in a recognition model's feature representation space, all
while avoiding visible changes. Our techniques work
under a wide range of assumptions and provide \dave{100\%} protection against
widely used, state-of-the-art models deployed by Microsoft, Amazon and
Face++. 

\revise{Like most privacy enhancing tools and technologies, Fawkes can also
  be used by malicious bad actors. For example, criminals
  could use Fawkes to hide their identity from agencies
  that rely on third-party facial recognition systems like Clearview.ai. We
  believe Fawkes will have the biggest impact on those using public images to
  build unauthorized facial recognition models and less so on agencies with
  legal access to facial images such as federal agencies or law enforcement. 
  We leave more detailed exploration of the tradeoff between user privacy and
  authorized use to future work.}

Protecting content using cloaks faces the inherent challenge of being {\em
  future-proof}, \dave{since} any technique we use to cloak images today might
be overcome by a workaround in some future date, which would render
previously protected images vulnerable. While we are under no illusion that
this proposed system is itself future-proof, we believe it is an important
and necessary first step in the development of user-centric privacy tools to
resist unauthorized machine learning models. We hope that followup work in
this space will lead to long-term protection mechanisms that
prevent the mining of personal content for user tracking and
classification. 

\secspace
\section*{Acknowledgments}
We thank our shepherd David Evans and anonymous reviewers for their
constructive feedback. This work is supported in part by NSF grants
CNS-1949650, CNS-1923778, CNS-1705042, and by the DARPA GARD program.  Any
opinions, findings, and conclusions or recommendations expressed in this
material are those of the authors and do not necessarily reflect the views of
any funding agencies.


{
 \small
 \bibliographystyle{acm}
 \bibliography{zhao,cloak}
}

\end{document}